%% file: main.tex
\pdfoutput=1
\documentclass[nouppercase,12pt]{ifmbe}
\title{ 
Diffusion MRI preprocessing affects ADC estimation and automatic PI-RADS v2.1 classification in bi-parametric prostate MRI}
\makeatletter
  \renewcommand{\ifmbe@processauthors}{%
    \def\ifmbe@author{%
      Christos Kanakis$^{1,3}$,
      Mathias Perslev$^{3}$,
      Tim Schakel$^{2}$,
      Silvia Ingala$^{3,4,5}$,
      Akshay Pai$^{3}$,
      Dennis Klomp$^{1}$,
      Chantal M.W. Tax$^{1}$%
    }%
  }
  \renewcommand{\ifmbe@processaffiliations}{%
    \def\ifmbe@affiliation{%
      \begin{minipage}{\textwidth}\centering\itshape\footnotesize
      $^{1}$Center for Image Sciences, University Medical Center Utrecht, Utrecht, The Netherlands\\
      $^{2}$Department of Radiotherapy, University Medical Center Utrecht, Utrecht, The Netherlands\\
      $^{3}$Cerebriu, Copenhagen, Denmark\\
      $^{4}$Department of Diagnostic Radiology, Copenhagen University Hospital Rigshospitalet, Copenhagen, Denmark\\
      $^{5}$Department of Diagnostic Radiology, Copenhagen University Hospital Herlev and Gentofte, Herlev, Denmark
      \end{minipage}%
    }%
  }
  \makeatother

\usepackage[sorting=none]{biblatex}
\usepackage{amsmath}
\usepackage{hyperref}
\usepackage{comment}
\usepackage{graphicx}
\usepackage{float}
\usepackage{booktabs}

\usepackage{caption}
\begin{document}

\pagestyle{plain}
\setlength{\footskip}{50pt}

\maketitle

\input{sections/00-abstract}

\input{sections/01-introduction}
\input{sections/02-materials-and-methods}

\input{sections/03-Results}
\input{sections/04-discussion}

\input{sections/05-Conclusion}

\input{sections/06-Acknowledgements}
\renewcommand*{\bibfont}{\footnotesize}
\printbibliography

\clearpage
\addcontentsline{toc}{section}{Supplementary Material} 

\end{document}

%% file: sections/00-abstract.tex
\begin{abstract}

\textit{Introduction}: Diffusion-weighted imaging (DWI) is acquired as part of bi-parametric and multi-parametric prostate MRI, but suffers from multiple artifacts that degrade downstream quantitative and diagnostic performance. While DWI preprocessing is more standard in brain imaging, its adoption in prostate imaging remains limited and lacks standardized pipelines. This study investigated the effect of different DWI preprocessing strategies on apparent diffusion coefficient (ADC) estimation and automatic Prostate Imaging Reporting and Data System (PI-RADS) classification performance.

\textit{Methods}: 268 cases were derived from the fastMRI prostate cohort by sequentially applying preprocessing steps: denoising, Gibbs-ringing correction, and diffeomorphic registration for susceptibility distortion correction, creating differently preprocessed datasets. First, ADC maps were compared using both linear least squares (LLS) and iteratively-weighted LLS (IWLLS) estimation. Next, a 3-class deep learning classifier based on a DenseNet architecture was trained to predict PI-RADS scores from multi-channel MRI inputs.

\textit{Results}: ADC analysis revealed statistically significant differences across preprocessing pipelines, with LLS and IWLLS estimation producing numerically equivalent ADC maps. Linear relationships between ADC values were preserved across most datasets (Pearson Correlation Coefficient, PCC~$\sim$0.99), while distortion correction realigned the DWI to the T2w anatomy and altered ADC values accordingly (PCC~$\sim$0.90). Automatic classification showed the best area under the receiver operating characteristic curve (AUROC) and sensitivity for high-risk PI-RADS classes in the fully processed dataset (Dataset 5). False negative (FN) analysis revealed that Dataset 5 produced the least overconfident incorrect predictions on high-risk classes, a desirable property for clinical triage.
\textit{Conclusion}: DWI preprocessing, particularly the inclusion of distortion correction, enhances both the quantitative quality of ADC maps and the predictive power of deep learning models for PI-RADS classification. These findings support the need for optimized DWI preprocessing pipelines in prostate MRI and highlight their potential to improve diagnostic accuracy and support automated triaging of high-risk prostate cancer.
\end{abstract}


%% file: sections/01-introduction.tex
\small
\section{Introduction}
Prostate cancer (PCa) is the most common malignancy affecting the male population. Magnetic Resonance Imaging (MRI) is currently the cornerstone of diagnostic and prognostic assessment of PCa, offering the possibility of risk stratification \cite{padhani2024key} and facilitating the selection of biopsy candidates \cite{launer2024contemporary} in certain populations. The Prostate Imaging Reporting and Data System (PI-RADS) v2.1 \cite{barrett2019pi} guides clinical decisions by assigning lesion scores from 1 to 5, reflecting the likelihood of clinically significant PCa (csPCa). PI-RADS scores of 1 or 2 are considered low risk, PI-RADS 3 is assigned to indeterminate cases and PI-RADS 4 or 5 are considered high risk and warrant further investigation. 

However, workflows can vary significantly among different institutions\cite{nalavenkata2025variation}: Multiparametric (mp)-MRI protocols include acquisition of T2-weighted (T2w) images, diffusion-weighted images (DWI) and dynamic contrast-enhanced (DCE) images using gadolinium. While some centers employ full mp-MRI with immediate contrast administration, others prefer bi-parametric (bp) MRI protocols that initially omit dynamic contrast-enhanced (DCE) imaging. In these cases, contrast is administered selectively after reviewing the initial bp-MRI results. These variations arise from differences in institutional expertise, resource availability, and local clinical guidelines, contributing to heterogeneous practices across healthcare systems. A major limitation in current practice is the significant inter-rater variability of PI-RADS interpretation \cite{taya2024perspectives}, which reduces diagnostic consistency and reliability. According to European Urology guidelines, patients with low-risk PI-RADS 1 or 2 scores may avoid biopsy, whereas higher scores typically prompt more aggressive investigation. However, inter-rater variability in PI-RADS scoring further exacerbates discrepancies, potentially impacting diagnostic confidence. 

DWI and the derived Apparent Diffusion Coefficient (ADC) maps are particularly critical for PI-RADS assessment \cite{barrett2019pi}. However, DWI is susceptible to artifacts caused by the strong gradient pulses used to encode water molecule displacements and the fast image readout strategy commonly employed \cite{le2006artifacts}. High \textit{b}-value images ($>1000$ s/mm²), essential for lesion detection \cite{godley2018accuracy}, often suffer from poor signal-to-noise ratio (SNR) and geometric distortions \cite{maier2022prostate}. In addition, contrast-to-noise ratio (CNR) optimized \textit{b}-values typically used for visual assessment are generally higher than those optimized for ADC accuracy \cite{fennessy2023quantitative,molendowska2024diffusion}. The Prostate Imaging Quality (PI-QUAL) v2 scoring system \cite{de2024pi} underscores the importance of high image quality (adequate CNR and SNR), for reducing inter-rater variability and ensuring reliable PI-RADS assignment. Low-quality DWI can result in an increased number of indeterminate PI-RADS 3 classifications, leading to unnecessary biopsies and additional contrast-enhanced imaging \cite{karanasios2022prostate}.

While in other fields such as brain imaging, efforts to compare and standardize preprocessing workflows for DWI data have been established \cite{veraart2022data}, no comparable consensus exists for prostate DWI. This lack of standardized preprocessing contributes to inconsistencies in image quality and diagnostic outcomes. Furthermore, there is currently no consensus supporting either bp-MRI or mp-MRI as the preferred approach \cite{guljavs2023dynamic}. Although bp-MRI has shown non-inferiority to mp-MRI in highly specialized settings, mp-MRI can offer additional insights for ambiguous or small lesions \cite{palumbo2020biparametric}. In particular, the added value of DCE remains debatable, as it has shown limited ability to clarify PI-RADS 3 lesions; quantitative ADC measurements alone can assist in upgrading them to PI-RADS 4 \cite{tavakoli2023contribution}. Collectively, these limitations — including variability in clinical workflows, inter-rater inconsistency, and the artifact-prone nature of DWI — highlight the urgent need for a more robust, standardized, and patient-specific approach to prostate MRI interpretation.

In recent years, Computer Aided Diagnosis (CAD) has emerged as a promising avenue to improve consistency and efficiency in prostate MRI interpretation, with a growing number of studies investigating its role in PCa management \cite{goldenberg2019new,van2021systematic}. Among CAD-based approaches, deep learning (DL)-driven PI-RADS classification has gained traction as a means to mitigate inter-rater variability. Although several studies have explored DL-based PI-RADS classification \cite{sanford2020deep,schelb2019classification}, none to date have systematically evaluated how DWI preprocessing pipelines affect DL performance. Given the increasing evidence that CAD-based methods can achieve diagnostic accuracy comparable to mp-MRI without the need for DCE imaging \cite{stanzione2019detection}, this study hypothesized that DWI preprocessing can improve the performance of DL-based PI-RADS classification on bp-MRI, thus effectively reducing the reliance on DCE scans.

To test this hypothesis, this study evaluated the impact of DWI preprocessing on: (a) ADC map estimation and (b) automated PI-RADS classification performance.

Confirmation of these benefits could support more tailored imaging pathways — such as selective contrast administration for PI-RADS 3 lesions and direct referral of PI-RADS 4/5 lesions for biopsy  — ultimately reducing unnecessary imaging and improving clinical efficiency.

%% file: sections/02-materials-and-methods.tex
\section{Materials and Methods}

\captionsetup[figure]{font=small}
\begin{figure*}[!t] 
    \centering
    \includegraphics[width=0.85\textwidth]{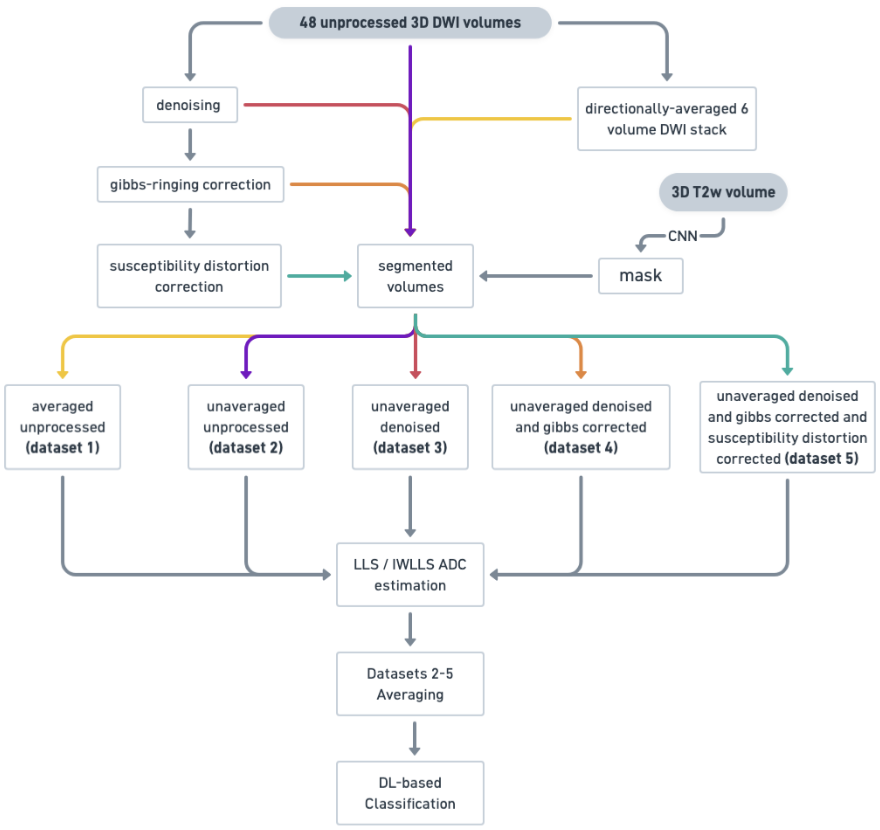}
    \caption{Technical and computational parts of this study's proposed workflow. Five differently preprocessed datasets were generated additively: Dataset 1 (directionally averaged, unprocessed) and Dataset 2 (unaveraged, unprocessed) served as baselines, while Datasets 3-5 added MPPCA denoising, Gibbs-ringing correction, and susceptibility distortion correction in sequence. All datasets were segmented and ADC maps were estimated within the prostate ROI using both LLS and IWLLS regression. The five datasets and the resulting ADC maps fed two downstream analyses: ADC quantification,
  and DL-based 3-class PI-RADS classification}
    \label{fig:1}
\end{figure*}

\subsection{Data} \label{data}

For the purposes of this study, 268 cases were obtained from the NYU fastMRI initiative database \cite{zbontar1811fastmri,knoll2020fastmri,tibrewala2024fastmri}. This dataset is the only publicly available prostate dataset that provides the raw k-space data, along with the necessary code to reconstruct the data. This uniquely enabled the assessment of DWI preprocessing on raw non-averaged data. For each case, 2-dimensional (2D) axial T2w turbo spin echo (TSE) [TR (s): 3.5-7.2, TE (ms): 100, in-plane resolution (mm): 0.56x0.56, Slice thickness (mm): 3, field of view (FOV) (mm): 180x180] and echo planar imaging (EPI)-DWI [TR (s): 5.0-7.3, TE (ms): 77, in-plane resolution (mm): 2.0x2.0, Slice thickness (mm): 3, FOV (mm): 200x200] sequences were acquired. For the latter, diffusion-sensitizing gradients were applied in three directions with \textit{b}-values of 50~s/mm\textsuperscript{2} and 1000~s/mm\textsuperscript{2}, with 4 and 12 averages respectively. 

T2w reconstruction included GeneRalized Autocalibration Partially Parallel Acquisition (GRAPPA) \cite{griswold2002generalized} and DWI reconstruction included EPI gridding, GRAPPA and an SNR-optimizing coil combination \cite{roemer1990nmr}. The dataset files stored the directionally averaged images for each \textit{b}-value, along with the k-space data. However, for the purposes of this study, the available reconstruction code was modified to output the individual DW images before averaging. This resulted in a single 3D axial T2w image and 48 DW images per subject — 12 acquired at a \textit{b}-value of 50~s/mm\textsuperscript{2} (4 per orthogonal direction) and 36 at a \textit{b}-value of 1000~s/mm\textsuperscript{2} (12 per orthogonal direction). 
T2w data were resampled to the DWI space to facilitate downstream tasks like registration (\ref{preproc}) and classification (\ref{class}). Figure \ref{fig:1} displays the technical and computational parts of this study's proposed workflow.
Furthermore, the dataset included slicewise PI-RADS v2.1 annotations of both T2w and DWI sequences, which were used in a DL-based classification task (\ref{class}).

\subsection{DWI Preprocessing} \label{preproc}

DWI data suffer from a wide variety of artifacts that are inherent to the application of diffusion-sensitizing gradients during scanning. 
The preprocessing in this work is focused on improving SNR, Gibbs-ringing, and susceptibility-induced distortion artifacts. 

DW images inherently have low SNR primarily caused by diffusion sensitizing gradients \cite{dikaios2014noise} and long TE, and denoising has become an increasingly adopted step to improve DWI quality. In this study, the Marcenko-Pastur Principal Component Analysis (MPPCA)-based denoising was used. It applies an orthogonal linear transformation to reduce dimensionality while preserving variance and removing components classified as noise \cite{veraart2016diffusion},\cite{marchenko1967distribution}. A patch radius of 2 voxels was used, therefore defining a 5x5x5 voxel sliding window.

Gibbs-ringing arises from the inverse Fourier transformation of k-space data during reconstruction, where high-frequency components are truncated due to acquisition time limitations. This truncation induces signal oscillations near regions of sharp intensity transitions—such as tissue or anatomical boundaries — manifesting as Gibbs-ringing artifacts \cite{veraart2016gibbs}. To mitigate this effect, a local sub-voxel shift suppression algorithm was applied  \cite{kellner2016gibbs}.

Although EPI-based sequences allow for rapid acquisition of diffusion images, they are highly susceptible to \( B_0 \) field inhomogeneities that cause geometric distortions along the phase-encoding (PE) direction \cite{jezzard1995correction}. In prostate imaging, this effect is exacerbated by the proximity to the air-filled rectum, which introduces large susceptibility gradients at the air--tissue interface \cite{mazaheri2013image}.  
In the absence of pairs of \( b_0 \) acquisitions with reversed PE to estimate a field map, this study adopted registration of the directionally averaged \( b_{50}\) image to the T2w image, with the use of a 3D symmetric diffeomorphic registration (SDR) algorithm \cite{avants2008symmetric}, modified to constrain the deformation along the PE axis (anterior-posterior (AP)). The generated displacement field is then applied to all DWI images and the inverse displacement field is stored for adjusting the T2W-derived mask used for later comparison (\ref{seg}). Geometric fidelity was assessed within the dilated prostate mask by comparing the b50 image to the T2w reference before and after correction, using the normalized gradient field (NGF) \cite{haber2006intensity} and mutual information (MI) \cite{maes1997multimodality} metrics, with a sign test on per-case differences. Field validity was checked via the Jacobian determinant of the displacement field, whose positivity indicates a smooth, invertible (diffeomorphic) deformation with no tissue folding.

Five differently processed datasets were generated for comparison. Dataset 1 consisted of averaged DWI data for each \textit{b}-value and direction ( \( b_{50x}\), \( b_{50y}\), \( b_{50z}\), \( b_{1000x}\), \( b_{1000y}\), \( b_{1000z}\) ) with no further preprocessing applied. Dataset 2 comprised the raw unaveraged unprocessed DWIs, individually reconstructed from k-space data. For Datasets 3-5, the preprocessing steps were applied additively — in the order mentioned above (denoising, Gibbs-ringing correction and susceptibility distortion correction) — on the 48 unaveraged DW scans (Dataset 2).

\subsection{Prostate Segmentation} \label{seg}  

The prostate gland was segmented from the T2w image to focus on the relevant region of interest (ROI) and ignore irrelevant anatomical structures in downstream analyses. An in-house implemented and evaluated 3D UNet-based \cite{ronneberger2015u} Convolutional Neural Network (CNN) was used for segmentation. The model, with [32,64,128,256] channels in each layer respectively, was trained on 6 publicly available datasets (1256 cases) including T2w images and manual annotations of the prostate gland. 15\% of the cases was excluded from training for independent evaluation, with the model achieving an average Dice score of 0.903, a value comparable to state-of-the-art frameworks \cite{rodrigues2023comparative}. Inference was performed on the fastMRI dataset T2w scans. A random subset of the resulting segmentations was visually inspected by a non-radiologist researcher with three years of experience in prostate MRI to screen for gross failures (e.g., bladder or rectum inclusion, missing prostate). No cases were excluded based on this screening. The computed segmentations could directly be used on Dataset 5 as DWIs were coregistered with the T2w image. For Datasets 1-4, the prostate mask was warped back to the DWI space using the reverse displacement field from DWI-to-T2 registration.

\subsection{Comparisons Between DWI Preprocessing pipelines}
The effect of preprocessing was first assessed by investigating differences in ADC quantification between datasets. Next, differences were evaluated on a clinically relevant PI-RADS score classification task using machine learning. 

\subsubsection{ADC quantification}
  Per-slice mean ADC was computed by averaging voxels inside the prostate segmentation mask, restricted to slices containing the prostate, yielding a slicewise mean-ADC distribution per dataset. Pairwise comparisons between datasets were performed on these distributions: the Wilcoxon signed-rank test assessed paired-slice differences, Pearson's correlation coefficient (PCC) quantified linear agreement, and Cohen's d \cite{diener2010cohen} and Wasserstein distance \cite{panaretos2019statistical} were calculated as effect-size and distributional measures. The same slicewise mean-ADC values were also stratified by PI-RADS class to examine per-score trends (median $\pm$ IQR/2).

\subsubsection{Automatic PI-RADS Classification} \label{class}

This experiment isolated preprocessing as the sole variable under investigation. The chosen architecture, hyperparameters, and training protocol were grounded in established prostate-MRI classification approaches \cite{yildirim2022deep,liu2021textured,youn2021detection,sanford2020deep} and held fixed across the four preprocessed datasets, allowing performance differences to be attributed to preprocessing rather than to architectural or hyperparameter optimization. Per-dataset hyperparameter search was not feasible given the cohort size (n=268), and would have conflated preprocessing effect with dataset-specific tuning effort; the composite-dataset training cutoff (\ref{class}) provides a partial mitigation by deriving a stable, cross-dataset training endpoint.

\begin{figure} [t]
    \centering
    \includegraphics[width=0.5\textwidth]{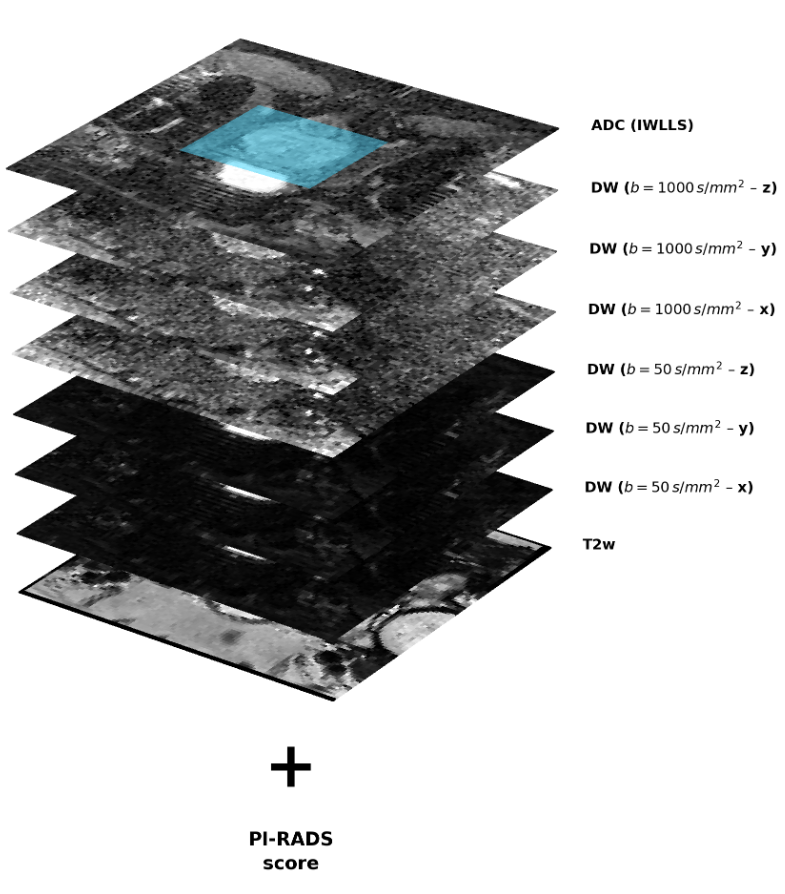}
    \caption{Visual representation of the model's expected input: A stack of 2D-slices derived from the T2w image, the averaged DW images and the ADC map, overlaid by the prostate mask's bounding box (in blue)}
    \label{fig:1.5}
\end{figure}

\textit{\textbf{Data preparation}}: Each axial slice was categorized into one of three classes based on its PI-RADS ground truth score as assessed on the DWI series: class 1 includes PI-RADS 1 and 2 (unlikely cancer), class 2 includes PI-RADS 3 (indeterminate, contrast enhanced MRI referral), and class 3 includes PI-RADS 4 and 5 (likely cancer, direct biopsy referral). As mentioned in \ref{data}, grouping was performed based on the initial dataset labels.
A fully automated deep learning-based classification framework was deployed using 2D slice-level inputs. To facilitate direct comparison between all datasets, DW volumes for Datasets 2-5 were averaged per \textit{b}-value and orthogonal direction to match Dataset 1 in the number of input channels, effectively rendering Datasets 1 and 2 identical for this specific task (averaged unprocessed data). Model input comprised 6 2D DW slices (one per \textit{b}-value and orthogonal direction), one 2D T2w slice, and one 2D ADC map slice, resulting in an 8-channel input. Slice intensities outside a filled bounding box around the segmented prostate were zeroed to focus the network on the prostate region while preserving immediate peri-prostatic context. Only ADC maps estimated using the IWLLS method were used in the classification task. Figure \ref{fig:1.5} provides a visual representation of the input.

\textbf{\textit{Model configuration:}} The classification model employed a 2D DenseNet architecture \cite{huang2017densely} with 64 initial features and a block configuration of [6, 12, 24, 16] filters. The model was trained using the Adam optimizer \cite{diederik2014adam} with a learning rate of $10^{-5}$, and the Focal Loss function \cite{lin2017focal}. Hyperparameters were chosen from literature defaults for medical-imaging DL classification, verified during initial training on the composite dataset (data-distribution-agnostic across preprocessing 
variants), and then held fixed across all four per-dataset runs. No specific preprocessed dataset served as a tuning anchor. As slicewise statistics across the dataset revealed that 93.31\% of slices belonged to class 1, extensive data augmentation \cite{kebaili2023deep} was incorporated to mitigate imbalance. Additionally, per-batch sampling was adapted depending on each class's recall, to boost exposure for classes with lower recall scores during training. All models were implemented in PyTorch 2.7.0 with CUDA 12.6 support \cite{imambi2021pytorch}, and trained on an NVIDIA GeForce RTX 3090 GPU with 24 GB memory. 

\captionsetup[figure]{font=small}
\begin{figure*}[!t] 
    \centering
    \includegraphics[width=\textwidth]{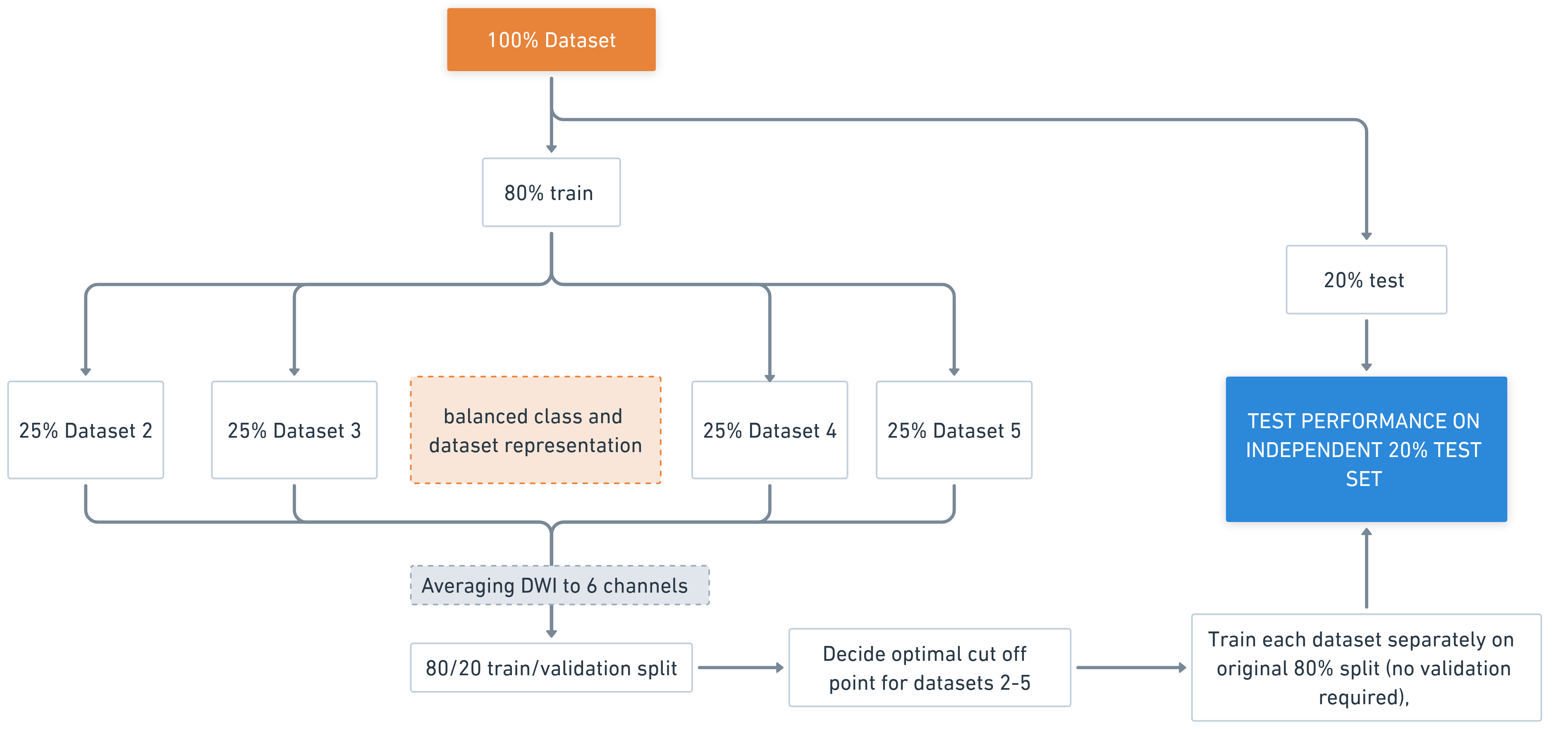}
    \caption{Configuration for DL-based classification experiments.}
    \label{fig:2}
    
\end{figure*}

\textbf{\textit{Experiment setup:}} With the available cohort size (n=268), per-dataset validation curves were too noisy for reliable training-cutoff selection. To obtain a stable signal on which to select a reliable training cut-off, a special model was trained on a composite dataset: 20\% of the cases were randomly held out as an independent test set with balanced class representation, and the remaining 80\% from all four preprocessed datasets (Datasets 2--5) were pooled into a single composite training set (25\% from each, with balanced PI-RADS representation), . A model was then trained on this composite set until convergence was clearly obtained based on the loss curves, after which a cut-off was chosen that minimized the validation loss, using an 80/20 train/validation split that was also balanced in both PI-RADS class and dataset representation. The EMA-smoothed composite validation curve was used to identify the epoch beyond which no further improvement was observed (figure \ref{fig:2}). This cutoff was then applied uniformly when training a separate model on each individual dataset, removing the need for per-dataset - and prohibitively unstable - validation sets.

\textbf{\textit{Evaluation:}} Slicewise per-class probabilities on the held-out test set were obtained by combining test-time augmentations \cite{shanmugam2020and} with an ensemble of the last 10 epochs before the training cutoff; patient-level probabilities were the mean across each patient's annotated slices, with ground truth defined as the maximum PI-RADS class per patient. Mean-pooling was
  preferred over maximum-pooling, which inflated patient-level false positives by assigning a high-risk label whenever a single slice produced an elevated class-2 or class-3 probability. Discriminative performance was quantified by one-vs-rest AUROC \cite{scikit-learn} at both slice and patient levels. Threshold-dependent metrics (sensitivity, specificity, precision) were derived via a repeated split-half procedure: at each of 1,000 iterations, the test set was randomly partitioned into class-balanced subsets A and B; class-specific thresholds were selected on A by grid search to maximize the macro-averaged F1 score \cite{grandini2020metrics} and applied to B to obtain patient-level predictions, with final estimates reported as mean $\pm$ standard deviation across iterations.
Pairwise Mann-Whitney U tests \cite{nachar2008mann} were performed between datasets on the per-iteration metric distributions for each class and metric. For patient-level false-negative (FN) predictions on the clinically critical classes (PI-RADS 3 and 4--5), the model's confidence in its (incorrect) predicted class was characterized via four softmax-derived metrics (maximum probability, margin, entropy, and variance), with lower maximum probability and margin and higher entropy and variance interpreted as reduced overconfidence in false predictions.

\captionsetup[figure]{font=small}
\begin{figure*}[!t] 
    \centering
    \includegraphics[width=\textwidth]{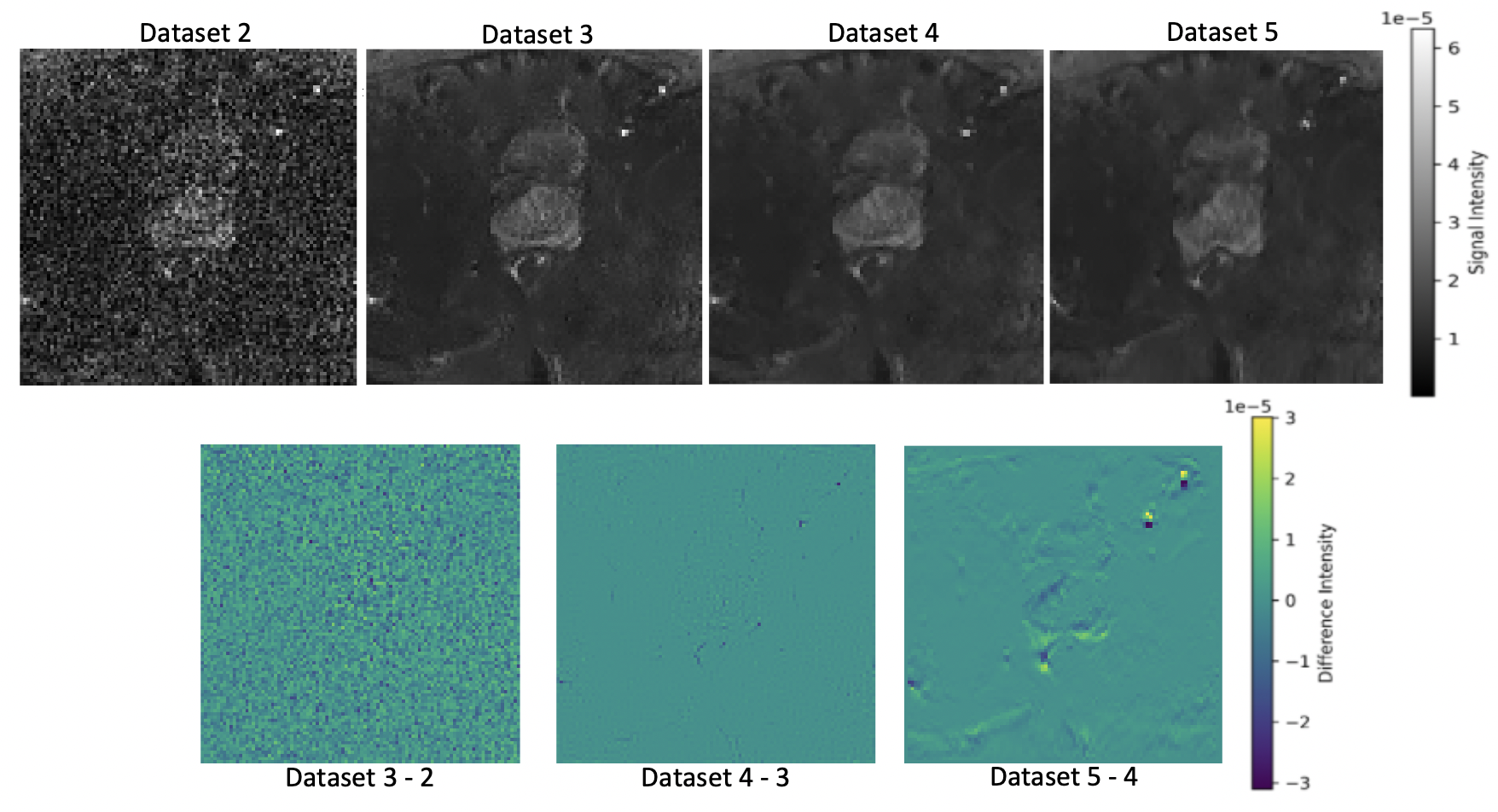}
    \caption{(up) an axial DW slice with \textit{b}-value 1000~s/mm\textsuperscript{2} across all preprocessing steps
             \newline(down): visualization of intensity changes after each preprocessing step }
    \label{fig:3}
\end{figure*}

%% file: sections/03-Results.tex
\section{Results}

In this study, the impact of DWI preprocessing on both ADC estimation and PI-RADS classification performance was evaluated. We first present results from slicewise ADC analyses across all preprocessing pipelines, followed by classification metrics from the deep learning framework. 

\captionsetup[figure]{font=small}
\begin{figure*}[!t] 
    \centering
    \includegraphics[width=0.85\textwidth]{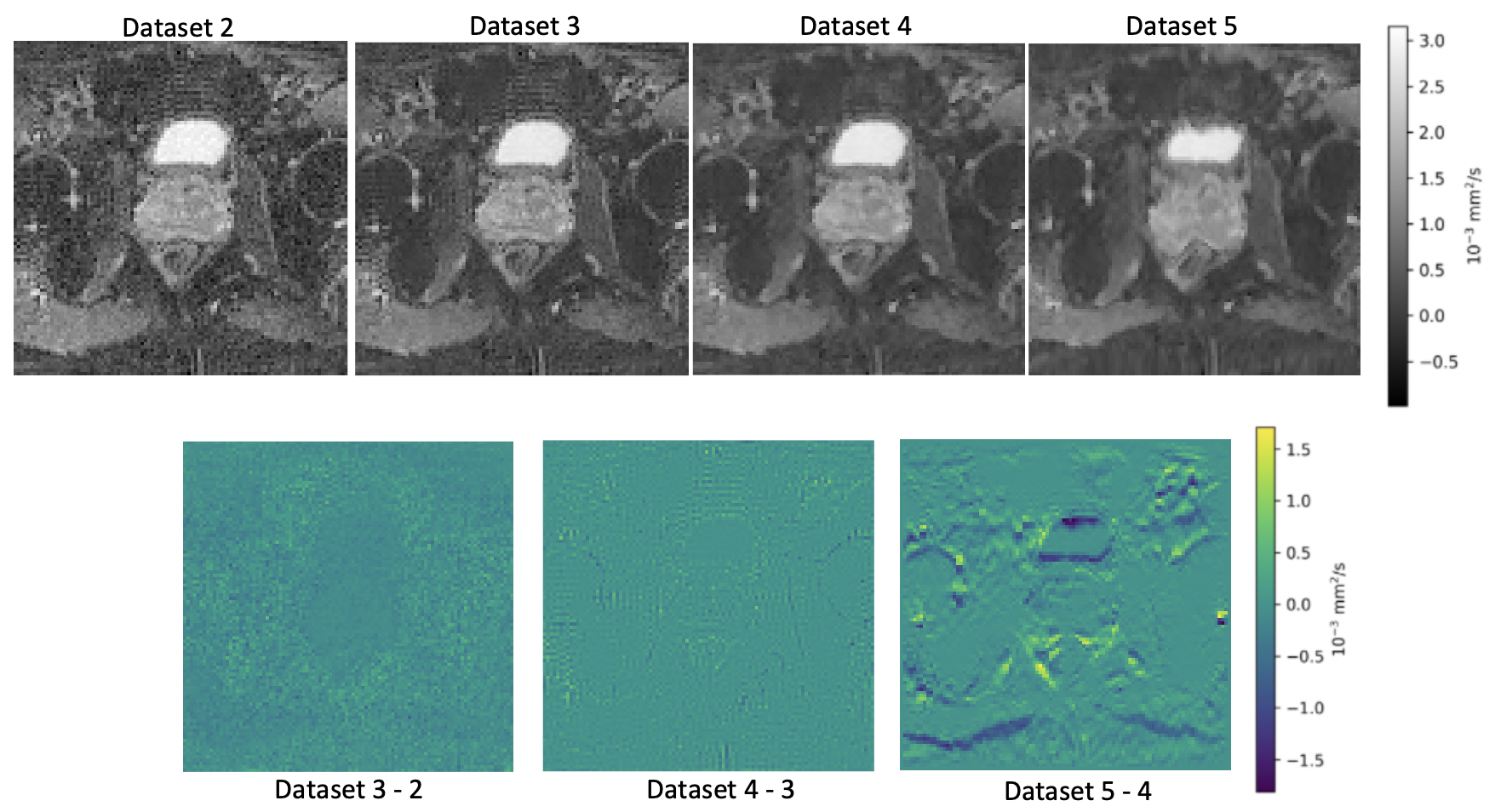}
    \caption{(up) an axial IWLLS-estimated ADC slice across all preprocessing steps 
             \newline(down): visualization of intensity changes after each preprocessing step }
    \label{fig:4}
\end{figure*}

\subsection{DWI preprocessing}
Figure \ref{fig:3} demonstrates an axial DW slice with \textit{b}-value 1000~s/mm\textsuperscript{2} across all preprocessing steps, along with a visualization of intensity changes per step. The following pattern, observed across all cases, emerges: Denoising affects the image in a uniform manner, by removing random noise on the entire DWI slice. Gibbs-ringing correction causes subtle intensity changes in certain areas. Susceptibility distortion correction appears to cause the most prominent changes in intensity values, especially across the Peripheral Zone (PZ) of the prostate gland. Within the dilated prostate mask, correction improved b50-to-T2w agreement: NGF alignment increased in 74\% of cases (198/268; sign test p<0.001) and MI in 79\% (213/268). The field remained diffeomorphic within the gland in 267 of 268 cases (positive Jacobian) and was strictly confined to the PE axis.

\subsection{ADC quantification}

Figure \ref{fig:4} demonstrates an axial IWLLS-estimated ADC slice across all preprocessing steps, along with a visualization of intensity changes per step. LLS and IWLLS estimation methods produced numerically equivalent ADC maps across all datasets: the absolute difference was on the order of $10^{-12}$ mm\textsuperscript{2}/s, with Cohen's d below 0.2 and Wasserstein distance below $0.001 \times 10^{-3}$ mm\textsuperscript{2}/s, confirming that the choice of estimation algorithm does not affect quantitative ADC values in this protocol. All subsequent analyses therefore focus on IWLLS-estimated maps. Slicewise PCC values revealed strong linear agreement across preprocessing configurations, consistently exceeding 0.89 \cite{schober2018correlation}. All pairwise comparisons between datasets were statistically significant (Wilcoxon signed-rank test, \textit{p} $\ll 0.001$); given the large sample size, effect sizes (Cohen's d and Wasserstein distance, table \ref{tab:1}) are more informative for interpreting between-dataset differences.

\captionsetup[figure]{font=small}
\begin{figure*}[t] 
    \centering
    \includegraphics[width=0.85\textwidth]{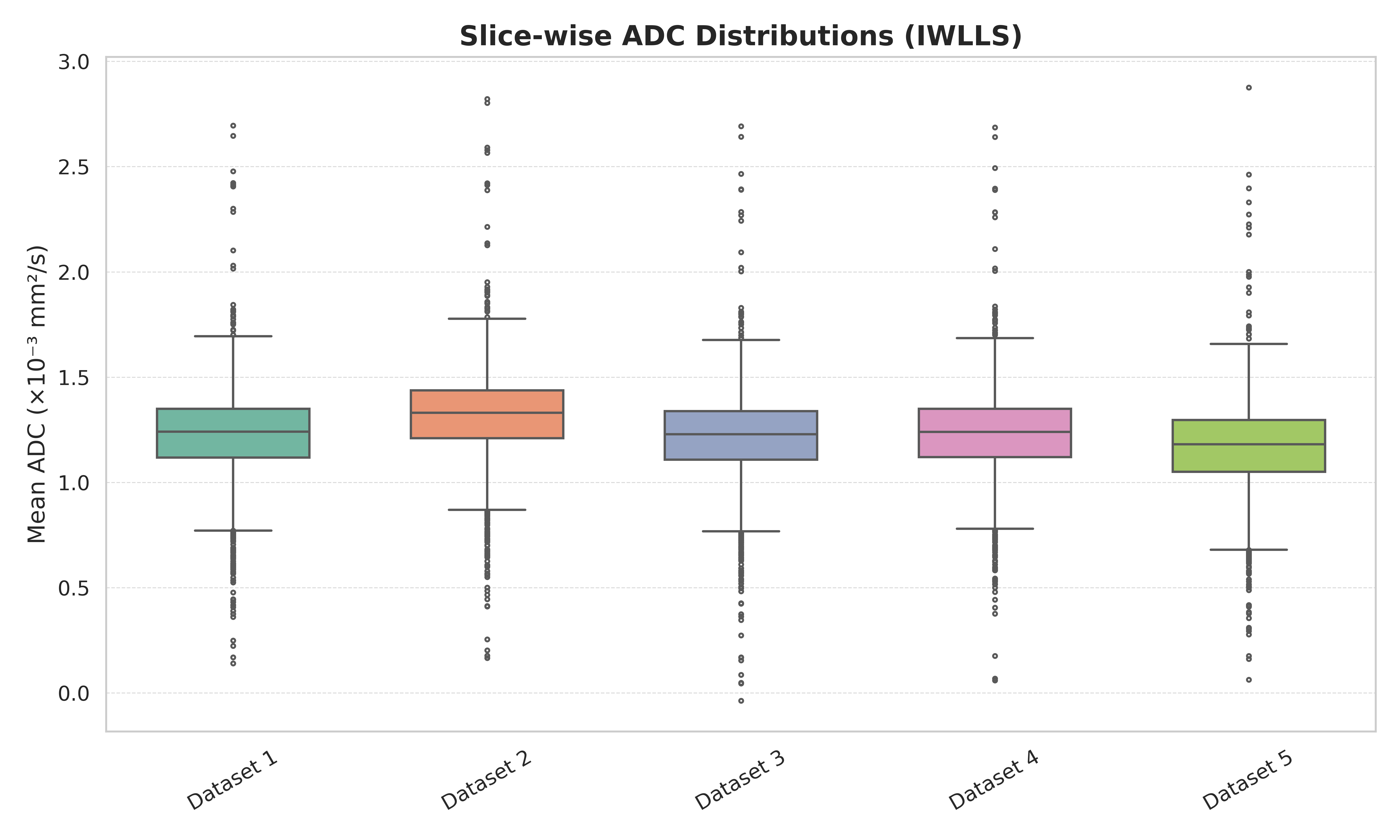}
    \caption{Slicewise mean-ADC distributions across all datasets for IWLLS}
    \label{fig:5}
\end{figure*}

\begin{figure*}[!t] 
    \centering
    \includegraphics[width=0.85\textwidth]{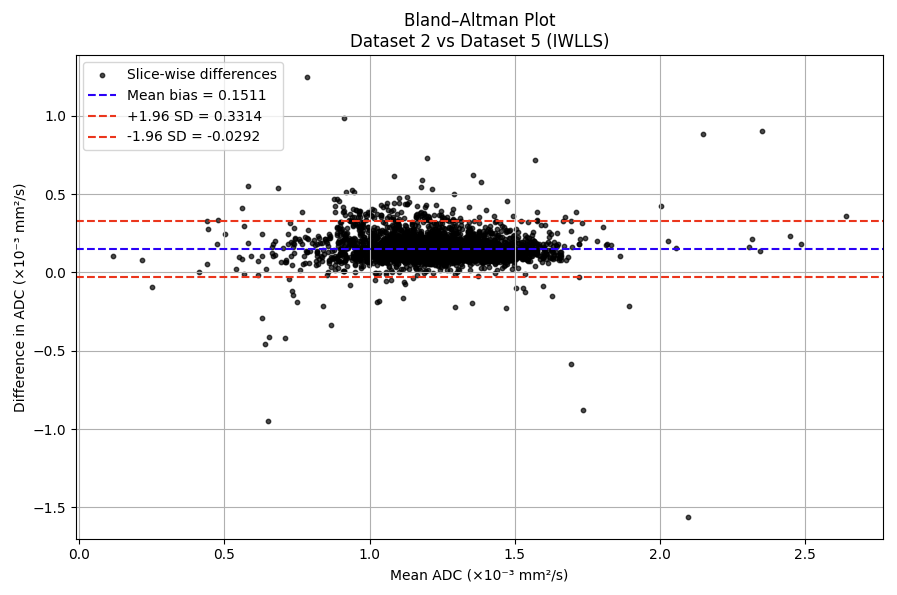}
    \caption{Bland-Altman plot of the slicewise differences between Datasets 2 and 5.}
    \label{fig:6}
\end{figure*}

Figure \ref{fig:5} demonstrates the slicewise mean-ADC distributions inside the prostate gland across all datasets for IWLLS. Dataset 2 exhibited the highest median ADC values compared to the rest. Corresponding Cohen's d and Wasserstein distance values are shown in table \ref{tab:1}.  Datasets 2 and 3 showed the highest d value, while Datasets 2 and 5 showed the highest Wasserstein distance. A Bland-Altman plot of the differences in slicewise ADC values between Datasets 2 and 5 can be observed in figure \ref{fig:6}, with the majority of the differences fluctuating between 0 and $0.4 \times 10^{-3}$ mm\textsuperscript{2}/s.
\captionsetup[figure]{font=small}

\begin{table*}[!t]
  \centering
  \small
  \begin{tabular}{@{}l ccccc@{}}
  \toprule
   & \textbf{Dataset 1} & \textbf{Dataset 2} & \textbf{Dataset 3} & \textbf{Dataset 4} & \textbf{Dataset 5} \\
  \midrule
  \textbf{Dataset 1} & ---   & 0.089             & 0.012 & 0.002 & 0.062 \\
  \textbf{Dataset 2} & 4.364 & ---               & 0.101 & 0.088 & \textbf{0.151} \\
  \textbf{Dataset 3} & 0.571 & \textbf{5.170}    & ---   & 0.013 & 0.051 \\
  \textbf{Dataset 4} & 0.037 & 3.384             & 0.450 & ---   & 0.063 \\
  \textbf{Dataset 5} & 0.676 & 1.642             & 0.543 & 0.719 & ---   \\
  \bottomrule
  \end{tabular}
  \caption{Pairwise distances between IWLLS-estimated slicewise mean-ADC distributions. Lower-left triangle: Cohen's $d$.
  Upper-right triangle: Wasserstein distance ($\times 10^{-3}$ mm\textsuperscript{2}/s). Bold values indicate the maximum within
  each triangle.}
  \label{tab:1}
  \end{table*}

\begin{figure*}[!t] 
    \centering
    \includegraphics[width=0.85\textwidth]{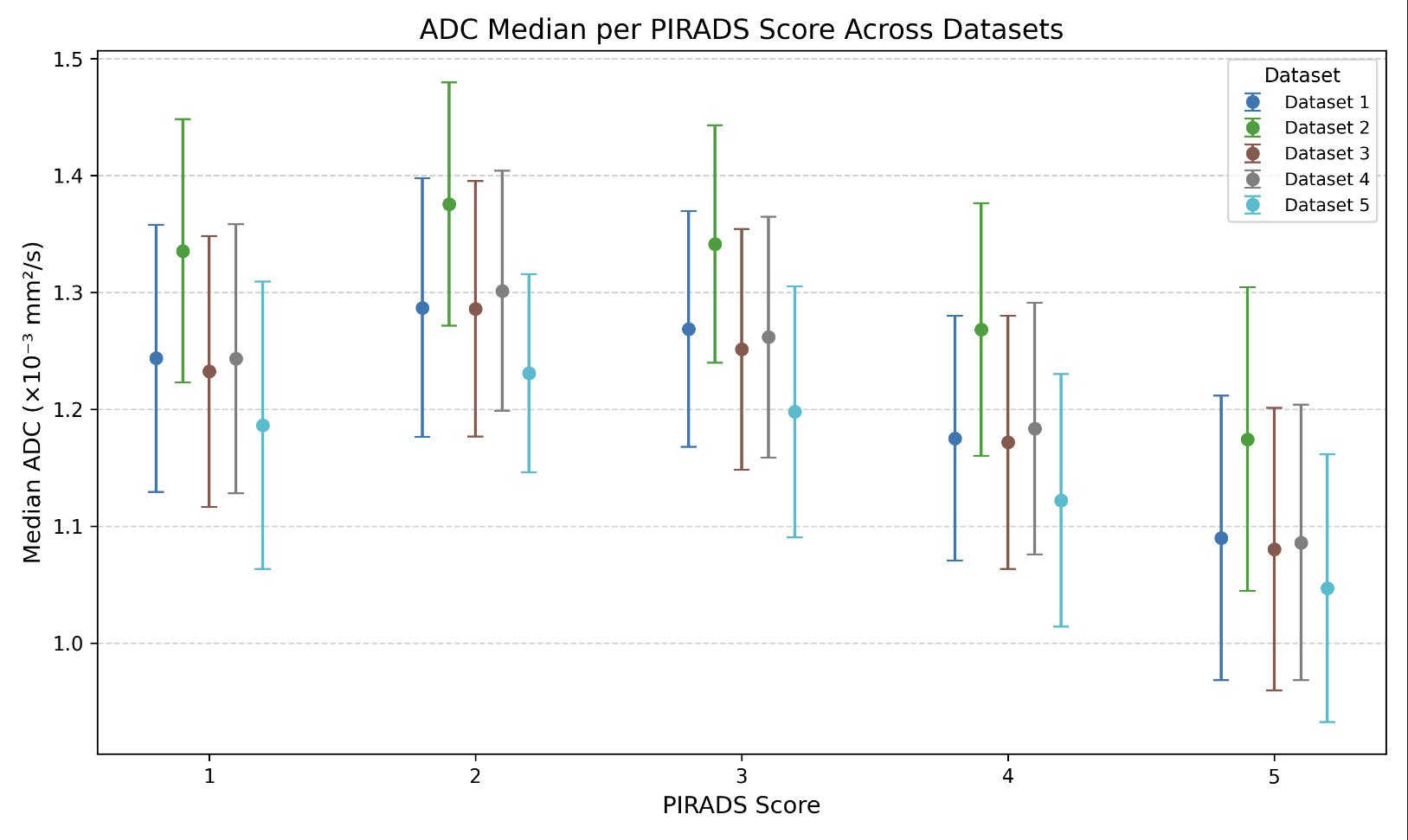}
    \caption{Median ADC values for the slicewise distribution per PI-RADS group for all IWLLS-estimated datasets. Error bars indicate ± ½ IQR}
    \label{fig:7}
\end{figure*}

We further investigated the relationship between ADC values and PI-RADS scores by comparing the slicewise mean-ADC distributions for PI-RADS 1-5 for all IWLLS-estimated ADC maps (figure \ref{fig:7}). The same overall trend was observed across all datasets: median values of the slicewise mean-ADC distribution for PI-RADS 1 slices were lower compared to PI-RADS 2. Median values of the slicewise mean-ADC distribution for PI-RADS 2-5 demonstrated a steady decrease when moving towards higher PI-RADS scores.  Statistical analysis of the distributions revealed significant differences across all methods for every PI-RADS score. No correction for multiple comparisons was applied, as all raw \textit{p}-values were orders of magnitude below any reasonable significance threshold, and correction would not alter any conclusions.

\captionsetup[figure]{font=small}

\subsection{Automatic PI-RADS Classification}

\begin{figure*}[!t] 
    \centering
    \includegraphics[width=\textwidth]{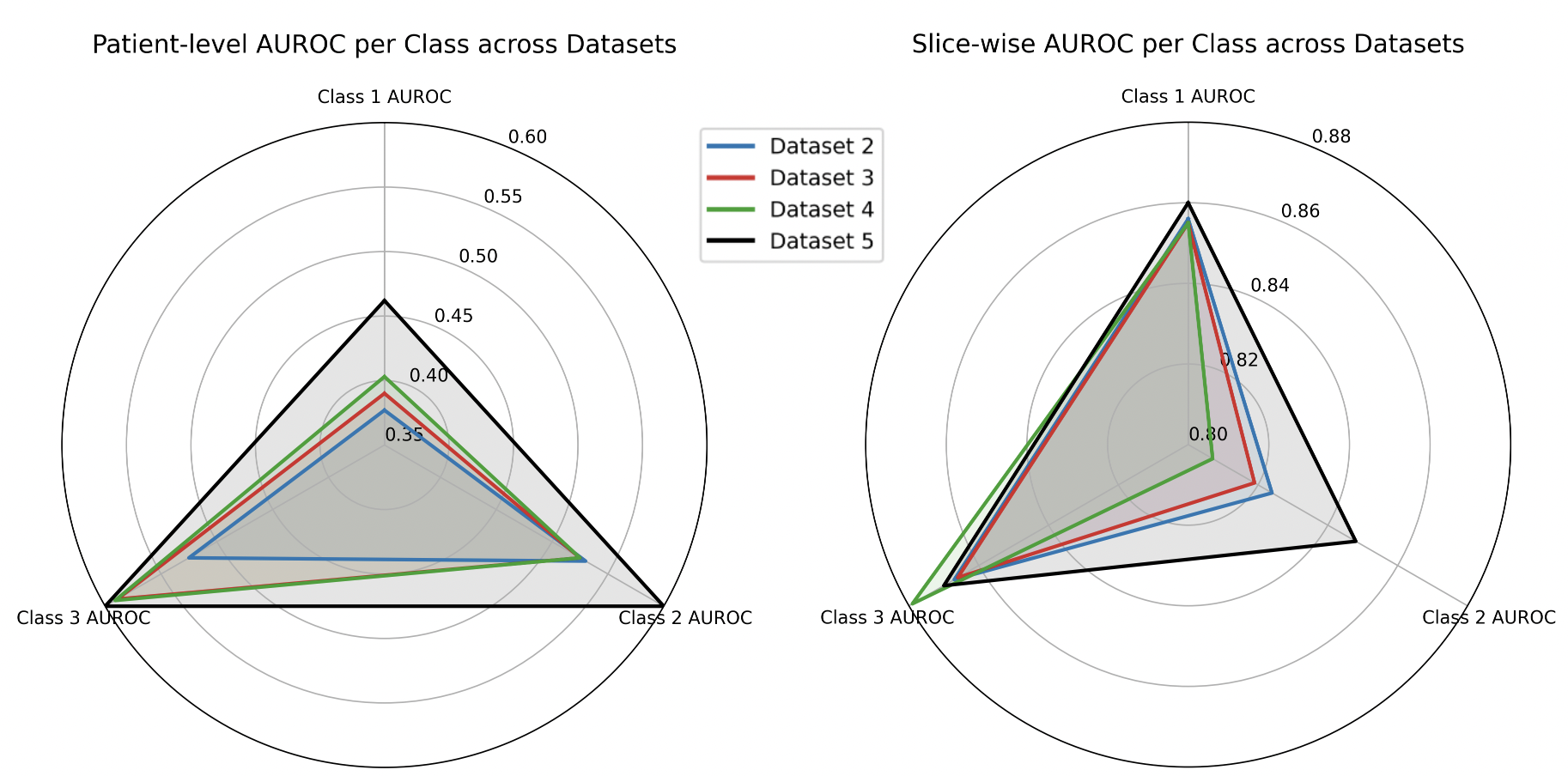}
    \caption{(a) Slicewise and (b) patient-level AUROC values per class across datasets. Dataset 2: averaged unprocessed, Dataset 3: averaged denoised, Dataset 4: averaged denoised + Gibbs corrected, Dataset 5: averaged denoised + Gibbs corrected + susceptibility distortion corrected}
    \label{fig:8}
\end{figure*}

\begin{figure*}[!t] 
    \centering
    \includegraphics[width=0.85\textwidth]{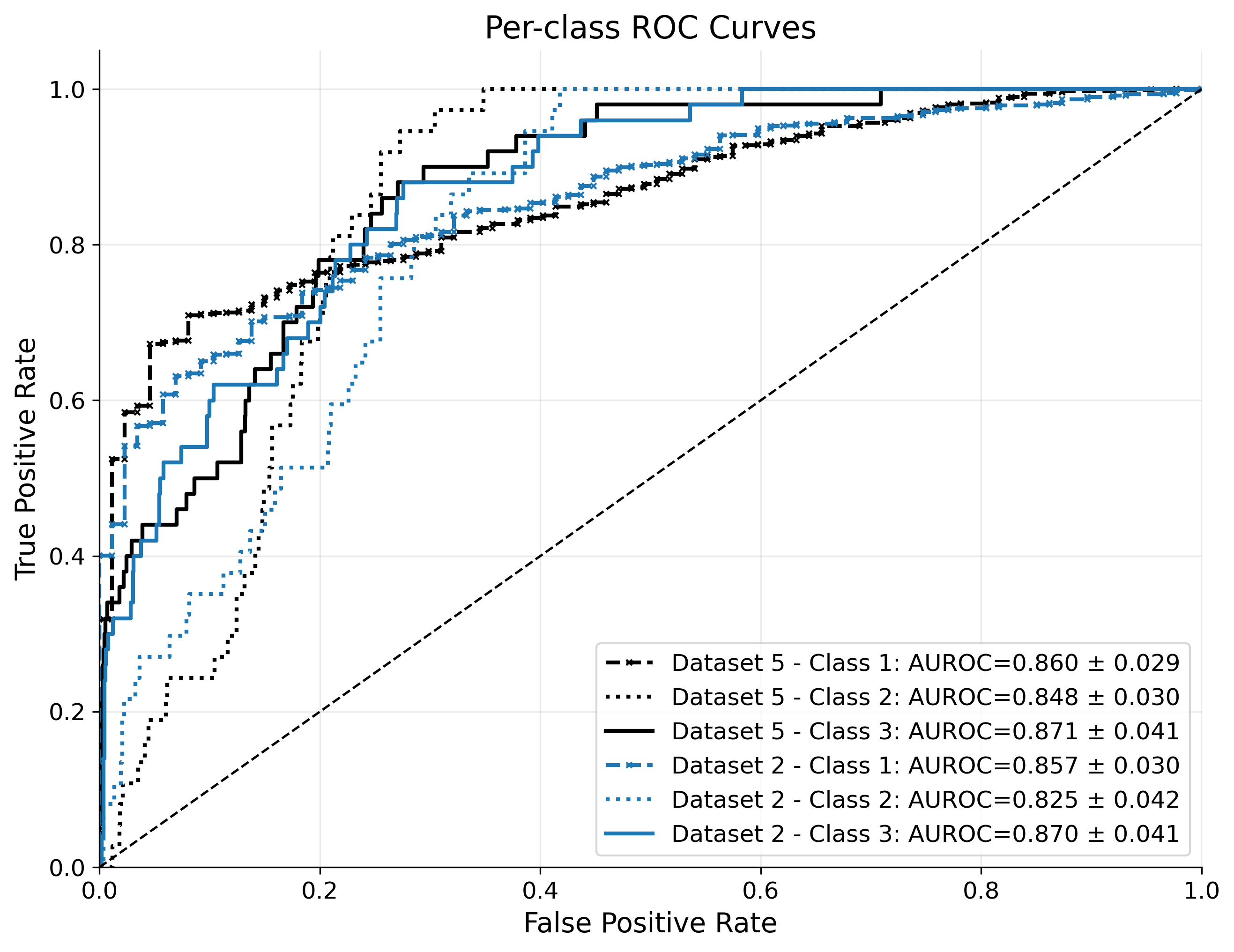}
    \caption{Per-class slicewise ROC curves for Datasets 2 and 5}
    \label{fig:9}
\end{figure*}

The optimal cut-off point was specified at 400 epochs, with no significant decrease in loss, or increase in per-class AUROC values after that point. Model weights for each dataset from epochs 391-400 were used for inference on the test set based on the evaluation approach mentioned in \ref{class}. Figure \ref{fig:8}(a) shows the slicewise AUROC values per class for all 4 datasets, with Dataset 5 (fully processed data) demonstrating superior performance across classes 1 and 2 and slightly inferior performance compared to Dataset 4, for class 3. Slicewise ROC curves for Datasets 2 and 5 are displayed in figure \ref{fig:9}. Focusing on clinically relevant cases (classes 2 and 3) and threshold-dependent metrics on a patient level, Dataset 5 demonstrated elevated sensitivity for both classes with values of 0.439 and 0.312 respectively. Dataset 5 also demonstrated slightly superior class 1 precision (0.392) and competitive class 2 and 3 precision (0.310 and 0.473 respectively). Detailed metrics can be seen in table \ref{tab:2}. Patient-level AUROC values demonstrated the superiority of Dataset 5 across all 3 classes, as seen in figure \ref{fig:8}(b). Mann-Whitney U tests showed statistically significant differences (\textit{p}~$<$~0.05) between almost all threshold-dependent metrics distributions (sensitivity, specificity, precision) across classes and datasets. Patient-level FN analysis for classes 2 and 3 revealed the following:  For class 2, the model trained on Dataset 5 was the least overconfident in its incorrect predictions, with the highest entropy (0.723), the lowest maximum softmax (0.747), the smallest margin (0.595), and the lowest variance (0.088). The same pattern held for class 3: Dataset 5 produced the highest entropy (0.716), the lowest maximum softmax (0.751), the lowest variance (0.089), and the second-lowest margin (0.592). In contrast, Dataset 2 produced the most overconfident incorrect predictions in both classes, with the lowest entropy and highest maximum softmax. These findings highlight Dataset 5 as the optimal choice for reducing overconfident FN errors in clinically important classes. 

\captionsetup[figure]{font=small}

\captionsetup[figure]{font=small}

\begin{table*}[!t]
  \centering
  \small
  \begin{tabular}{@{}l l cccc@{}}
  \toprule
  \textbf{Metric} & \textbf{Class} & \textbf{Dataset 2} & \textbf{Dataset 3} & \textbf{Dataset 4} & \textbf{Dataset 5} \\
  \midrule
  \multirow{3}{*}{\textbf{Sensitivity}}
    & 1 & \textbf{0.635 $\pm$ 0.255} & 0.371 $\pm$ 0.212          & 0.477 $\pm$ 0.279          & 0.381 $\pm$ 0.177 \\
    & 2 & 0.331 $\pm$ 0.192          & 0.360 $\pm$ 0.179          & 0.237 $\pm$ 0.199          & \textbf{0.439 $\pm$ 0.187} \\
    & 3 & 0.132 $\pm$ 0.098          & 0.367 $\pm$ 0.241          & 0.253 $\pm$ 0.181          & \textbf{0.312 $\pm$ 0.198} \\
  \addlinespace
  \multirow{3}{*}{\textbf{Specificity}}
    & 1 & 0.297 $\pm$ 0.169          & 0.437 $\pm$ 0.198          & 0.377 $\pm$ 0.245          & \textbf{0.568 $\pm$ 0.193} \\
    & 2 & \textbf{0.776 $\pm$ 0.178} & 0.747 $\pm$ 0.140          & 0.732 $\pm$ 0.211          & 0.638 $\pm$ 0.133 \\
    & 3 & \textbf{0.951 $\pm$ 0.084} & 0.813 $\pm$ 0.113          & 0.848 $\pm$ 0.140          & 0.837 $\pm$ 0.153 \\
  \addlinespace
  \multirow{3}{*}{\textbf{Precision}}
    & 1 & 0.369 $\pm$ 0.127          & 0.312 $\pm$ 0.112          & 0.308 $\pm$ 0.139          & \textbf{0.392 $\pm$ 0.134} \\
    & 2 & \textbf{0.391 $\pm$ 0.180} & 0.342 $\pm$ 0.156          & 0.247 $\pm$ 0.156          & 0.310 $\pm$ 0.123 \\
    & 3 & \textbf{0.527 $\pm$ 0.378} & 0.392 $\pm$ 0.227          & 0.417 $\pm$ 0.265          & 0.473 $\pm$ 0.269 \\
  \bottomrule
  \end{tabular}
  \caption{Patient-level classification metrics (mean $\pm$ standard deviation across 1{,}000 split-half iterations) per class
  and per preprocessed dataset, after grid-search threshold optimisation. Class~1: PI-RADS 1--2; class~2: PI-RADS 3; class~3:
  PI-RADS 4--5. Bold values indicate the best within each metric--class row.}
  \label{tab:2}
  \end{table*}

%% file: sections/04-discussion.tex
\section{Discussion}

This study assessed the effect of prostate DWI preprocessing on two downstream tasks: ADC estimation and automated PI-RADS classification. Three preprocessing pipelines, two averaging strategies, and two ADC estimation algorithms were evaluated against these endpoints.

\subsection{DWI preprocessing}

Artifact correction is more commonly adopted for brain DWI\cite{tax2022s,veraart2022data}, yet a range of studies addresses artifact correction for prostate DWI. More specifically: i) eddy current-nulled, convex-optimised diffusion encoding (EN-CODE)~\cite{zhang2020prostate} has been shown to reduce eddy-current-induced distortion in prostate DWI while shortening TE relative to standard monopolar or twice-refocused bipolar gradient designs. The FSL eddy toolbox, initially developed for brain DWI \cite{andersson2016integrated}, has also been used \cite{lemberskiy2018characterization}. ii) DL-based denoising approaches \cite{kaye2020accelerating,han2023prostate,pfaff2024enhancing} have been proven to improve SNR ratio, especially for high \textit{b}-value images. Alternatively, the Marchenko-Pastur principal component analysis (MPPCA) algorithm has also been employed for denoising \cite{pfaff2024enhancing,zhang2024high}, showing promising results. It should be noted that all mentioned approaches require raw diffusion data (non-averaged) to be implemented. iii) Gibbs-ringing correction is part of several DWI preprocessing pipelines \cite{palombo2023joint,molendowska2024diffusion,grussu2021feasibility} that implement the algorithm also used in this study (see \ref{preproc}). iv) Motion-related artifacts are commonly reduced by fast EPI-based sequences \cite{tamada2021diffusion,ueno2022diffusion}. Alternatively, post-acquisition methods involve several registration techniques \cite{buerger2015comparing}, with shown superiority of the "fast elastic image registration" method for prostate DWI \cite{kabus2010fast}. v) Geometric and susceptibility-induced distortions in prostate DWI are mitigated through acquisition strategies such as reduced field-of-view and segmented EPI, which limit phase-encoding distortions \cite{rakowpenner2015prostate,bergen2020field,ciris2019accelerated}. Retrospective field map-based corrections using B0 mapping or reversed phase-encoding further improve spatial fidelity \cite{nketiah2018geometric,tong2019exploratory,syversen2018implementation}. More advanced model-based reconstructions integrate susceptibility field estimation directly into the reconstruction process, enabling refined geometric corrections and minimizing local displacement errors \cite{usman2020joint,usman2019model}.
Even though previous work addresses prostate DWI preprocessing, to our knowledge, no systematic comparison of how different pipelines affect an end task has been performed. Moreover, the pipelines mentioned above are not routinely implemented in clinical settings for PCa diagnosis and risk stratification.

Regarding other artifacts, for the fastMRI dataset, GRAPPA was employed to enable faster acquisition with minimal SNR loss and mitigate acquisition-time related artifacts such as signal drift \cite{vos2017importance}. Furthermore, ghosting artifacts were reduced through parallel imaging strategies \cite{larkman2007parallel}. Specifically, the fastMRI dataset used parallel imaging with an acceleration factor of R = 2 for the EPI-DWI scans and employed coil combinations to optimize SNR, addressing the inherent low SNR challenges of EPI-based sequences \cite{le2006artifacts}. In addition to SNR improvements, these sequences help suppress motion-related artifacts and improve anatomical alignment, supporting their established role in bp-MRI for prostate cancer \cite{jambor2017optimization}. However, EPI sequences remain  sensitive to magnetic field inhomogeneities, leading to geometric distortions that are particularly pronounced on 3T systems, such as the scanner used in this study \cite{mazaheri2013image}. A limitation of the current dataset is the absence of \( b_0 \) scans, which, while not essential for deep learning applications, preclude the use of established susceptibility correction methods that rely on reversed phase-encoding (PE) scans, such as FSL TOPUP \cite{andersson2003correct} or the DRBUDDI method from TORTOISE \cite{irfanoglu2015dr}. These so-called “blip-up” and “blip-down” scans have become more standard practice in recent susceptibility distortion correction approaches for brain DWI \cite{tax2022s}. Moreover, eddy current–induced artifacts in prostate DWI are typically addressed at acquisition through optimized gradient designs (e.g., EN-CODE) rather than retrospectively. This contrasts with brain DWI, where high SNR and tissue contrast make image-registration-based retrospective corrections such as FSL eddy reliable enough that protocols such as the Human Connectome Project forgo bipolar (eddy-cancelling) gradient designs in favour of more efficient monopolar encoding. In body applications, by contrast, the lower SNR and reduced intra-tissue contrast within the organ render registration-based EC correction less reliable and difficult to validate~\cite{molendowska2024diffusion}, motivating the focus on acquisition-side mitigation in the prostate. Given previous work and the current dataset's sequence design, this study prioritized the correction of noise, Gibbs-ringing, and magnetic susceptibility artifacts.
Although visual evaluation confirmed differences in image quality between preprocessing steps (figure \ref{fig:3}), downstream DL-based classification more accurately estimates the clinical value of those steps. A direct geometric assessment confirmed this: b50-to-T2w NGF and MI agreement within the prostate improved after correction (74\% and 79\% of cases) with the field remaining diffeomorphic in the gland, indicating improved rather than degraded alignment.

\subsection{ADC analysis}

In this study, the effect of DWI preprocessing on ADC estimation was investigated, which is important if absolute values are to be used in quantitative comparisons or clinical decision-making. The purpose was to establish whether preprocessing pipelines produce clinically meaningful differences in ADC quantification.
LLS and IWLLS estimation methods were found to be numerically equivalent for this protocol, with differences on the order of $10^{-12}$ mm\textsuperscript{2}/s — well below floating-point precision and clinically meaningless. This confirmed that for the given dataset, ADC estimation algorithm selection does not affect quantitative values, and all further discussion focuses on IWLLS-estimated maps. Slicewise PCC values indicated very strong linear relationships ($\sim 0.99$) between Datasets 1, 2, 3 and 4. This outcome is not only expected but desired, since processes like averaging, denoising and Gibbs-ringing correction should increase image quality while preserving anatomical contrast \cite{veraart2016denoising}. Dataset 5 showed a lower linear relationship with the rest ($\sim$ 0.90), reflecting that distortion correction deliberately re-aligned the DWI to the T2w anatomy, unlike the geometry-preserving steps; this geometric change propagated to the ADC maps rather than indicating loss of fidelity.  Based on Wasserstein distance, Datasets 2 and 5 were the most different in terms of slicewise mean-ADC distributions, which confirmed that the additive effect of the preprocessing pipelines has significant quantitative impact on the data. Datasets 1 and 4 had the most similar distributions, confirming that Gibbs-ringing correction focused only on anatomical boundaries \cite{veraart2016gibbs}, whereas denoising and distortion correction cause global quantitative changes. It should also be noted that the assumptions about the most similar and different ADC distributions were made based on Wasserstein distance and not Cohen's d, since those values can be more easily interpretable in a clinical context because they are in the same units as the data (ADC values in mm\textsuperscript{2}/s). I.e., a difference of $0.15 \times 10^{-3}$ mm\textsuperscript{2}/s (see \ref{tab:1}) between Datasets 2 and 5 can be considered clinically significant in a range of ADC values between 0.7 and 2.0 $\times 10^{-3}$ mm\textsuperscript{2}/s for the prostate gland \cite{thormer2012diagnostic}.
To further support the clinical context of the study, the behavior of ADC values within the prostate gland as a function of slicewise PI-RADS annotation was also investigated. From figure \ref{fig:7} it can be observed that the median value of the slicewise mean-ADC distribution consistently decreases for higher PI-RADS scores, for all datasets. This observation aligned with the presentation of lesions as low-signal areas on the ADC maps indicating restricted diffusion, therefore decreasing the mean ADC values \cite{manetta2019correlation}. However, this pattern is not consistent for PI-RADS 1 observations. Although the PI-RADS 1 median was expected to exceed the PI-RADS 2 median, the observed value was lower. This discrepancy is partially attributed to the fact that PI-RADS 1 slices make up almost 96\% of the total dataset and span the entire prostate gland, including the transition zone (TZ), which exhibits lower ADC values than the peripheral zone (PZ). The resulting zonal composition differences between PI-RADS 1 and PI-RADS 2 slices likely confound slicewise mean ADC comparisons. Additionally, the presence of conditions such as Benign Prostatic Hyperplasia (BPH) that mimic hypo intense regions, which have higher ADC values than lesions but lower values compared to healthy tissue \cite{emad2014diffusion}, further contributes to this pattern. Considering that evidence of BPH are identified in 50\% of men between 50-60 years old \cite{vuichoud2015benign} and the fact that this particular dataset has a mean subject age of 66 ± 8 years \cite{tibrewala2024fastmri}, it is possible that the downwards trend in mean slicewise PI-RADS 1 values can be attributed to this.

\subsection{Automatic PI-RADS Classification}

This study employed an automatic PI-RADS classification task to assess whether DWI preprocessing enhances the predictive performance of DL-based models. The PI-RADS scores were grouped into 3 clinically relevant classes (PI-RADS 1-2: class 1, PI-RADS 3: class 2, PI-RADS 4-5: class 3), aligning with clinical decision points for contrast administration and biopsy referral. The objective was to compare relative performance across preprocessing pipelines rather than to compete with state-of-the-art architectures \cite{yildirim2022deep,liu2021textured,youn2021detection,sanford2020deep}. Unlike most DL-based studies that frame csPCa detection as a binary problem \cite{yoo2019prostate} or use 3D architectures \cite{winkel2020autonomous}, this study employed a slicewise 2D framework to leverage the dataset's unique slicewise PI-RADS labels (similarly to \cite{schelb2019classification}), effectively increasing the sample pool and statistical power. To ensure fair comparison, all datasets were averaged to a 6-DWI-channel input, rendering Datasets 1 and 2 identical and limiting comparison to Datasets 2-5. Since preprocessing was applied to the unaveraged DWI volumes, differences in model performance reflect the impact of the preprocessing strategies. Due to the limited dataset size, conventional train/validation/test splits were not feasible, and a composite dataset strategy was employed to derive the training cutoff via EMA-smoothed validation loss. While per-dataset validation was not feasible, the four datasets share the same patient cohort, labels, architecture, and hyperparameters and differ only in image preprocessing; the composite-derived cutoff is therefore not anchored to any single preprocessing variant. The class distribution in the test set was balanced as much as possible to minimize evaluation bias \cite{kang2020balanced}.

Threshold-agnostic metrics (AUROC values) on both slice- and patient-level showcased the superiority of Dataset 5 (fully processed). On a slice level specifically, Dataset 5 demonstrated the highest discriminative ability for classes 1 and 2, whereas on a patient level, Dataset 5 demonstrated the highest discriminative ability in all classes. It should be noted that even though the test set is balanced across classes on a patient level, slice-level imbalance remains severe, since only a few slices per patient potentially belong to class 2 or 3. Although previous studies have reported that ROC curves are overly optimistic with regards to imbalanced datasets \cite{davis2006relationship}, it has recently been shown that they are an accurate representation of the models discriminative capabilities \cite{richardson2024receiver}. Threshold-dependent metrics on the other hand, are less clear as to which dataset performs best. Given the clinical context of the task, focus should be shifted on class 2 and 3 sensitivity. Improved sensitivity for high-risk cases reflects the model's ability to capture those cases, which is essential in PCa risk stratification. Dataset 5 performed best in these cases, thus reducing underdiagnosis of high-risk cases. Precision values, however, showed that Dataset 2 performed best in classes 2 and 3, while Dataset 5 performed best in class 1. Higher precision in classes 2 and 3 reduces false positives, resulting in fewer unnecessary biopsies. This trend reflects the established trade-off between detecting as many true cases as possible (sensitivity) and ensuring that positive predictions are correct (precision) \cite{buckland1994relationship}, which is central to real-world diagnostic triage. FN analysis, however, showed that Dataset 5 was least confident when missing class 2 or 3 cases, a behavior highly desirable in the given clinical context. This pattern aligns with the growing emphasis on probabilistic calibration in clinical deep-learning systems, where well-expressed uncertainty in erroneous predictions is a prerequisite for safe automated triage \cite{guo2017calibration}; the present results suggest that DWI preprocessing modulates calibration in addition to accuracy. Larger data availability, allowing for proper model optimization and threshold selection strategies could further enhance the behavior of Dataset 5 and increase classification performance across all classes. Although the repeated split-half procedure mitigates overfitting to the test set, it does not fully substitute for an independent validation set, and reported threshold-dependent metrics should be interpreted with this caveat. Given that fully processed data showed promising results with competitive (and often higher) performance against unprocessed data, enhancing the robustness of the preprocessing steps and optimizing model architecture and decision boundaries on a larger scale, could pave the way for a standardized DWI preprocessing protocol that enhances DL model capabilities and supports automated triaging of high-risk PCa patients. An open question for future work is whether preprocessing could help counter data scarcity — i.e., whether a smaller set of fully processed data could achieve comparable performance to a larger set of unprocessed data, effectively increasing the value of each acquired dataset.

\subsection{Strengths and limitations}

One of the main strengths of this study was that the unique access to raw k-space data from the fastMRI prostate dataset enabled preprocessing of individual, unaveraged DW images — a step that is not possible with standard DICOM-based clinical data — which in turn maximized the effectiveness of denoising and artifact correction algorithms. Moreover, the additive pipeline design (Datasets 1--5) allowed the effect of each preprocessing step to be isolated and evaluated independently. Additionally, the clinically grounded 3-class PI-RADS framework directly aligned with clinical decision points (observation, contrast referral, biopsy), ensuring that the evaluation metrics reflected real-world diagnostic utility. Finally, the FN uncertainty analysis provided insight beyond standard accuracy metrics, demonstrating that fully processed data reduces overconfident misclassifications in high-risk cases.

Regarding this study's limitations, all data originated from a single site and scanner (Siemens Skyra), limiting generalizability to other field strengths, vendors, and acquisition protocols. The absence of reversed phase-encoding acquisitions precluded the use of established susceptibility correction methods (e.g., FSL TOPUP); the PE-constrained diffeomorphic registration employed here, while effectively correcting geometric distortion, could not recover intensity distortion (signal pile-up and signal loss) inherent to susceptibility-induced EPI distortion. Future prospective studies should incorporate reversed phase-encoding acquisitions — requiring only approximately 60 seconds of additional scan time \cite{nketiah2018geometric} — to enable both geometric and intensity distortion correction. The dataset size (268 cases) precluded conventional train/validation/test splits, necessitating the composite dataset strategy described above. The severe class imbalance (96\% PI-RADS 1) reflected clinical reality but limited the statistical power of class 2 and 3 analyses. Slicewise PI-RADS labels, while enabling the 2D approach, could not fully capture the spatial extent of lesions that spanned multiple slices. Finally, a complementary radiological manual reader study can be considered to evaluate inter-rater variability and assess increase in reader confidence through preprocessing.

%% file: sections/05-Conclusion.tex
\section{Conclusion}

This study demonstrates that DWI preprocessing significantly affects ADC quantification and DL-based PI-RADS classification in prostate bp-MRI. While the choice of ADC estimation algorithm (LLS vs. IWLLS) had no clinically meaningful impact, sequential application of denoising, Gibbs-ringing correction, and susceptibility distortion correction produced quantitatively and clinically significant differences in ADC values. The fully preprocessed dataset (Dataset 5) achieved the highest AUROC and sensitivity for high-risk PI-RADS classes. Critically, when the model was wrong on these classes, it expressed the lowest confidence in the incorrect label, a desirable property for clinical triage. These findings support the adoption of (standardized) DWI preprocessing pipelines in prostate MRI and highlight the need for routine reversed phase-encoding acquisition to enable comprehensive distortion correction. Future work should validate these findings across multi-site, multi-vendor cohorts, investigate whether DWI preprocessing can effectively augment limited datasets to improve DL model performance, and quantify the clinical impact of preprocessing on radiological readability through a dedicated multi-reader assessment.

%% file: sections/06-Acknowledgements.tex
\section{Acknowledgements}
This project is supported by the Eurostars 3 Project "PRISMA" (Prostate Imaging  Solutions with MRI Automation); Project id ESTAR24203.
CT is supported by the Dutch Research Council (NWO) under a VIDI grant with file number 21299.
This work made use of the fastMRI Prostate dataset made publicly available by NYU Langone Health Center for Advanced Imaging Innovation and Research \cite{zbontar1811fastmri,knoll2020fastmri,tibrewala2024fastmri}.

%% file: references.bib
@article{padhani2024key,
  title={Key learning on the promise and limitations of MRI in prostate cancer screening},
  author={Padhani, Anwar R and Godtman, Rebecka A and Schoots, Ivo G},
  journal={European Radiology},
  volume={34},
  number={9},
  pages={6168--6174},
  year={2024},
  publisher={Springer}
}

@inproceedings{launer2024contemporary,
  title={A contemporary review: mpMRI in prostate cancer screening and diagnosis},
  author={Launer, Bryn M and Ellis, Taryn A and Scarpato, Kristen R},
  booktitle={Urologic Oncology: Seminars and Original Investigations},
  year={2024},
  organization={Elsevier}
}

@article{barrett2019pi,
  title={PI-RADS version 2.1: one small step for prostate MRI},
  author={Barrett, Tristan and Rajesh, Arumugam and Rosenkrantz, Andrew B and Choyke, Peter L and Turkbey, Baris},
  journal={Clinical radiology},
  volume={74},
  number={11},
  pages={841--852},
  year={2019},
  publisher={Elsevier}
}

@article{taya2024perspectives,
  title={Perspectives on technology: Prostate Imaging-Reporting and Data System (PI-RADS) interobserver variability},
  author={Taya, Michio and Behr, Spencer C and Westphalen, Antonio C},
  journal={BJU international},
  volume={134},
  number={4},
  pages={510--518},
  year={2024},
  publisher={Wiley Online Library}
}

@article{le2006artifacts,
  title={Artifacts and pitfalls in diffusion MRI},
  author={Le Bihan, Denis and Poupon, Cyril and Amadon, Alexis and Lethimonnier, Franck},
  journal={Journal of Magnetic Resonance Imaging: An Official Journal of the International Society for Magnetic Resonance in Medicine},
  volume={24},
  number={3},
  pages={478--488},
  year={2006},
  publisher={Wiley Online Library}
}

@inproceedings{veraart2022data,
  title={A data-driven variability assessment of brain diffusion MRI preprocessing pipelines},
  author={Veraart, Jelle and Christiaens, Daan and Dai, Erpeng and Edwards, Luke J and Golkov, Vladimir and Mohammadi, Siawoosh and Schilling, Kurt G and Aarabi, Mohammad Hadi and Ades-Aron, Benjamin and Adluru, Nagesh and others},
  booktitle={Proceedings of the Joint Annual ISMRM-ESMRMB Meeting},
  year={2022}
}

@article{maier2022prostate,
  title={Prostate cancer diffusion-weighted magnetic resonance imaging: does the choice of diffusion-weighting level matter?},
  author={Maier, Stephan E and Wallstr{\"o}m, Jonas and Langkilde, Fredrik and Johansson, Jens and Kuczera, Stefan and Hugosson, Jonas and Hellstr{\"o}m, Mikael},
  journal={Journal of Magnetic Resonance Imaging},
  volume={55},
  number={3},
  pages={842--853},
  year={2022},
  publisher={Wiley Online Library}
}

@article{fennessy2023quantitative,
  title={Quantitative diffusion MRI in prostate cancer: Image quality, what we can measure and how it improves clinical assessment},
  author={Fennessy, Fiona M and Maier, Stephan E},
  journal={European journal of radiology},
  volume={167},
  pages={111066},
  year={2023},
  publisher={Elsevier}
}

@article{de2024pi,
  title={PI-QUAL version 2: an update of a standardised scoring system for the assessment of image quality of prostate MRI},
  author={de Rooij, Maarten and Allen, Clare and Twilt, Jasper J and Thijssen, Linda CP and Asbach, Patrick and Barrett, Tristan and Brembilla, Giorgio and Emberton, Mark and Gupta, Rajan T and Haider, Masoom A and others},
  journal={European radiology},
  volume={34},
  number={11},
  pages={7068--7079},
  year={2024},
  publisher={Springer}
}

@article{karanasios2022prostate,
  title={Prostate MRI quality: clinical impact of the PI-QUAL score in prostate cancer diagnostic work-up},
  author={Karanasios, Eleftherios and Caglic, Iztok and Zawaideh, Jeries P and Barrett, Tristan},
  journal={The British Journal of Radiology},
  volume={95},
  number={1133},
  pages={20211372},
  year={2022},
  publisher={The British Institute of Radiology.}
}

@article{palumbo2020biparametric,
  title={Biparametric (bp) and multiparametric (mp) magnetic resonance imaging (MRI) approach to prostate cancer disease: a narrative review of current debate on dynamic contrast enhancement},
  author={Palumbo, Pierpaolo and Manetta, Rosa and Izzo, Antonio and Bruno, Federico and Arrigoni, Francesco and De Filippo, Massimo and Splendiani, Alessandra and Di Cesare, Ernesto and Masciocchi, Carlo and Barile, Antonio},
  journal={Gland Surgery},
  volume={9},
  number={6},
  pages={2235--2247},
  year={2020}
}

@article{guljavs2023dynamic,
  title={Dynamic Contrast-Enhanced Study in the mpMRI of the Prostate—Unnecessary or Underutilised? A Narrative Review},
  author={Gulja{\v{s}}, Silva and Dupan Krivdi{\'c}, Zdravka and Dre{\v{z}}njak Maduni{\'c}, Maja and {\v{S}}ambi{\'c} Penc, Mirela and Pavlovi{\'c}, Oliver and Krajina, Vinko and Pavokovi{\'c}, Deni and {\v{S}}mit Taka{\v{c}}, Petra and {\v{S}}tefan{\v{c}}i{\'c}, Marin and Salha, Tamer},
  journal={Diagnostics},
  volume={13},
  number={22},
  pages={3488},
  year={2023},
  publisher={MDPI}
}

@article{stanzione2019detection,
  title={Detection of extraprostatic extension of cancer on biparametric MRI combining texture analysis and machine learning: preliminary results},
  author={Stanzione, Arnaldo and Cuocolo, Renato and Cocozza, Sirio and Romeo, Valeria and Persico, Francesco and Fusco, Ferdinando and Longo, Nicola and Brunetti, Arturo and Imbriaco, Massimo},
  journal={Academic radiology},
  volume={26},
  number={10},
  pages={1338--1344},
  year={2019},
  publisher={Elsevier}
}

@article{tavakoli2023contribution,
  title={Contribution of dynamic contrast-enhanced and diffusion MRI to PI-RADS for detecting clinically significant prostate cancer},
  author={Tavakoli, Anoshirwan Andrej and Hielscher, Thomas and Badura, Patrick and G{\"o}rtz, Magdalena and Kuder, Tristan Anselm and Gnirs, Regula and Schwab, Constantin and Hohenfellner, Markus and Schlemmer, Heinz-Peter and Bonekamp, David},
  journal={Radiology},
  volume={306},
  number={1},
  pages={186--199},
  year={2023},
  publisher={Radiological Society of North America}
}

@article{zbontar1811fastmri,
  title={fastMRI: An open dataset and benchmarks for accelerated MRI},
  author={Zbontar, J and Knoll, F and Sriram, A and Murrell, T and Huang, Z and Muckley, MJ and Defazio, A and Stern, R and Johnson, P and Bruno, M and others},
  journal={arXiv preprint arXiv:1811.08839},
  year={2018}
}

@article{knoll2020fastmri,
  title={fastMRI: A publicly available raw k-space and DICOM dataset of knee images for accelerated MR image reconstruction using machine learning},
  author={Knoll, Florian and Zbontar, Jure and Sriram, Anuroop and Muckley, Matthew J and Bruno, Mary and Defazio, Aaron and Parente, Marc and Geras, Krzysztof J and Katsnelson, Joe and Chandarana, Hersh and others},
  journal={Radiology: Artificial Intelligence},
  volume={2},
  number={1},
  pages={e190007},
  year={2020},
  publisher={Radiological Society of North America}
}

@article{tibrewala2024fastmri,
  title={FastMRI Prostate: A public, biparametric MRI dataset to advance machine learning for prostate cancer imaging},
  author={Tibrewala, Radhika and Dutt, Tarun and Tong, Angela and Ginocchio, Luke and Lattanzi, Riccardo and Keerthivasan, Mahesh B and Baete, Steven H and Chopra, Sumit and Lui, Yvonne W and Sodickson, Daniel K and others},
  journal={Scientific Data},
  volume={11},
  number={1},
  pages={404},
  year={2024},
  publisher={Nature Publishing Group UK London}
}

@article{andersson2003correct,
  title={How to correct susceptibility distortions in spin-echo echo-planar images: application to diffusion tensor imaging},
  author={Andersson, Jesper LR and Skare, Stefan and Ashburner, John},
  journal={Neuroimage},
  volume={20},
  number={2},
  pages={870--888},
  year={2003},
  publisher={Elsevier}
}

@article{griswold2002generalized,
  title={Generalized autocalibrating partially parallel acquisitions (GRAPPA)},
  author={Griswold, Mark A and Jakob, Peter M and Heidemann, Robin M and Nittka, Mathias and Jellus, Vladimir and Wang, Jianmin and Kiefer, Berthold and Haase, Axel},
  journal={Magnetic Resonance in Medicine: An Official Journal of the International Society for Magnetic Resonance in Medicine},
  volume={47},
  number={6},
  pages={1202--1210},
  year={2002},
  publisher={Wiley Online Library}
}

@article{vos2017importance,
  title={The importance of correcting for signal drift in diffusion MRI},
  author={Vos, Sjoerd B and Tax, Chantal MW and Luijten, Peter R and Ourselin, Sebastien and Leemans, Alexander and Froeling, Martijn},
  journal={Magnetic resonance in medicine},
  volume={77},
  number={1},
  pages={285--299},
  year={2017},
  publisher={Wiley Online Library}
}

@article{larkman2007parallel,
  title={Parallel magnetic resonance imaging},
  author={Larkman, David J and Nunes, Rita G},
  journal={Physics in Medicine \& Biology},
  volume={52},
  number={7},
  pages={R15--R55},
  year={2007},
  publisher={IOP Publishing}
}

@article{dikaios2014noise,
  title={Noise estimation from averaged diffusion weighted images: Can unbiased quantitative decay parameters assist cancer evaluation?},
  author={Dikaios, Nikolaos and Punwani, Shonit and Hamy, Valentin and Purpura, Pierpaolo and Rice, Scott and Forster, Martin and Mendes, Ruheena and Taylor, Stuart and Atkinson, David},
  journal={Magnetic Resonance in Medicine},
  volume={71},
  number={6},
  pages={2105--2117},
  year={2014},
  publisher={Wiley Online Library}
}

@article{veraart2016diffusion,
  title={Diffusion MRI noise mapping using random matrix theory},
  author={Veraart, Jelle and Fieremans, Els and Novikov, Dmitry S},
  journal={Magnetic resonance in medicine},
  volume={76},
  number={5},
  pages={1582--1593},
  year={2016},
  publisher={Wiley Online Library}
}

@article{veraart2016gibbs,
  title={Gibbs ringing in diffusion MRI},
  author={Veraart, Jelle and Fieremans, Els and Jelescu, Ileana O and Knoll, Florian and Novikov, Dmitry S},
  journal={Magnetic resonance in medicine},
  volume={76},
  number={1},
  pages={301--314},
  year={2016},
  publisher={Wiley Online Library}
}

@article{kellner2016gibbs,
  title={Gibbs-ringing artifact removal based on local subvoxel-shifts},
  author={Kellner, Elias and Dhital, Bibek and Kiselev, Valerij G and Reisert, Marco},
  journal={Magnetic resonance in medicine},
  volume={76},
  number={5},
  pages={1574--1581},
  year={2016},
  publisher={Wiley Online Library}
}

@article{jezzard1995correction,
  title={Correction for geometric distortion in echo planar images from B0 field variations},
  author={Jezzard, Peter and Balaban, Robert S},
  journal={Magnetic resonance in medicine},
  volume={34},
  number={1},
  pages={65--73},
  year={1995},
  publisher={Wiley Online Library}
}

@article{tax2022s,
  title={What’s new and what’s next in diffusion MRI preprocessing},
  author={Tax, Chantal MW and Bastiani, Matteo and Veraart, Jelle and Garyfallidis, Eleftherios and Irfanoglu, M Okan},
  journal={NeuroImage},
  volume={249},
  pages={118830},
  year={2022},
  publisher={Elsevier}
}

@article{avants2008symmetric,
  title={Symmetric diffeomorphic image registration with cross-correlation: evaluating automated labeling of elderly and neurodegenerative brain},
  author={Avants, Brian B and Epstein, Charles L and Grossman, Murray and Gee, James C},
  journal={Medical image analysis},
  volume={12},
  number={1},
  pages={26--41},
  year={2008},
  publisher={Elsevier}
}

@inproceedings{ronneberger2015u,
  title={U-net: Convolutional networks for biomedical image segmentation},
  author={Ronneberger, Olaf and Fischer, Philipp and Brox, Thomas},
  booktitle={Medical image computing and computer-assisted intervention--MICCAI 2015: 18th international conference, Munich, Germany, October 5-9, 2015, proceedings, part III 18},
  pages={234--241},
  year={2015},
  organization={Springer}
}

@article{rodrigues2023comparative,
  title={A comparative study of automated deep learning segmentation models for prostate MRI},
  author={Rodrigues, Nuno M and Silva, Sara and Vanneschi, Leonardo and Papanikolaou, Nickolas},
  journal={Cancers},
  volume={15},
  number={5},
  pages={1467},
  year={2023},
  publisher={MDPI}
}

@inproceedings{huang2017densely,
  title={Densely connected convolutional networks},
  author={Huang, Gao and Liu, Zhuang and Van Der Maaten, Laurens and Weinberger, Kilian Q},
  booktitle={Proceedings of the IEEE conference on computer vision and pattern recognition},
  pages={4700--4708},
  year={2017}
}

@article{diederik2014adam,
  title={Adam: A method for stochastic optimization},
  author={Kingma, Diederik P. and Ba, Jimmy},
  journal={arXiv preprint arXiv:1412.6980},
  year={2014}
}

@article{imambi2021pytorch,
  title={PyTorch},
  author={Imambi, Sagar and Prakash, Kolla Bhanu and Kanagachidambaresan, GR},
  journal={Programming with TensorFlow: solution for edge computing applications},
  pages={87--104},
  year={2021},
  publisher={Springer}
}

@article{schober2018correlation,
  title={Correlation coefficients: appropriate use and interpretation},
  author={Schober, Patrick and Boer, Christa and Schwarte, Lothar A},
  journal={Anesthesia \& analgesia},
  volume={126},
  number={5},
  pages={1763--1768},
  year={2018},
  publisher={LWW}
}

@article{veraart2016denoising,
  title={Denoising of diffusion MRI using random matrix theory},
  author={Veraart, Jelle and Novikov, Dmitry S and Christiaens, Daan and Ades-Aron, Benjamin and Sijbers, Jan and Fieremans, Els},
  journal={Neuroimage},
  volume={142},
  pages={394--406},
  year={2016},
  publisher={Elsevier}
}

@article{emad2014diffusion,
  title={Diffusion-weighted MR imaging and ADC measurement in normal prostate, benign prostatic hyperplasia and prostate carcinoma},
  author={Emad-Eldin, Sally and Halim, Manal and Metwally, Lamiaa IA and Abdel-Aziz, Rania Mahmoud},
  journal={The Egyptian Journal of Radiology and Nuclear Medicine},
  volume={45},
  number={2},
  pages={535--542},
  year={2014},
  publisher={Elsevier}
}

@article{vuichoud2015benign,
  title={Benign prostatic hyperplasia: epidemiology, economics and evaluation},
  author={Vuichoud, Camille and Loughlin, Kevin R},
  journal={Can J Urol},
  volume={22},
  number={Suppl 1},
  pages={1--6},
  year={2015}
}

@article{sanford2020deep,
  title={Deep-learning-based artificial intelligence for PI-RADS classification to assist multiparametric prostate MRI interpretation: A development study},
  author={Sanford, Thomas and Harmon, Stephanie A and Turkbey, Evrim B and Kesani, Deepak and Tuncer, Sena and Madariaga, Manuel and Yang, Chris and Sackett, Jonathan and Mehralivand, Sherif and Yan, Pingkun and others},
  journal={Journal of Magnetic Resonance Imaging},
  volume={52},
  number={5},
  pages={1499--1507},
  year={2020},
  publisher={Wiley Online Library}
}

@article{diener2010cohen,
  title={Cohen's d},
  author={Diener, Marc J},
  journal={The Corsini encyclopedia of psychology},
  pages={1--1},
  year={2010},
  publisher={Wiley Online Library}
}

@article{panaretos2019statistical,
  title={Statistical aspects of Wasserstein distances},
  author={Panaretos, Victor M and Zemel, Yoav},
  journal={Annual review of statistics and its application},
  volume={6},
  number={1},
  pages={405--431},
  year={2019},
  publisher={Annual Reviews}
}

@article{thormer2012diagnostic,
  title={Diagnostic value of ADC in patients with prostate cancer: influence of the choice of b values},
  author={Th{\"o}rmer, Gregor and Otto, Josephin and Reiss-Zimmermann, Martin and Seiwerts, Matthias and Moche, Michael and Garnov, Nikita and Franz, Toni and Do, Minh and Stolzenburg, Jens-Uwe and Horn, Lars-Christian and others},
  journal={European radiology},
  volume={22},
  pages={1820--1828},
  year={2012},
  publisher={Springer}
}

@article{roemer1990nmr,
  title={The NMR phased array},
  author={Roemer, Peter B and Edelstein, William A and Hayes, Cecil E and Souza, Steven P and Mueller, Otward M},
  journal={Magnetic resonance in medicine},
  volume={16},
  number={2},
  pages={192--225},
  year={1990},
  publisher={Wiley Online Library}
}

@article{manetta2019correlation,
  title={Correlation between ADC values and Gleason score in evaluation of prostate cancer: multicentre experience and review of the literature},
  author={Manetta, Rosa and Palumbo, Pierpaolo and Gianneramo, Camilla and Bruno, Federico and Arrigoni, Francesco and Natella, Raffaele and Maggialetti, Nicola and Agostini, Andrea and Giovagnoni, Andrea and Di Cesare, Ernesto and others},
  journal={Gland surgery},
  volume={8},
  number={Suppl 3},
  pages={S216--S222},
  year={2019}
}

@article{kebaili2023deep,
  title={Deep learning approaches for data augmentation in medical imaging: a review},
  author={Kebaili, Aghiles and Lapuyade-Lahorgue, J{\'e}r{\^o}me and Ruan, Su},
  journal={Journal of imaging},
  volume={9},
  number={4},
  pages={81},
  year={2023},
  publisher={MDPI}
}

@article{nalavenkata2025variation,
  title={Variation in Prostate Magnetic Resonance Imaging Performance: Data from the Prostate Biopsy Collaborative Group},
  author={Nalavenkata, Sunny B and Vertosick, Emily and Briganti, Alberto and Ahmed, Hashim and Eldred-Evans, David and Gordon, Steven and Raghallaigh, Holly and Gratzke, Christian and O’Callaghan, Michael and Liss, Michael and others},
  journal={European Urology Oncology},
  year={2025},
  publisher={Elsevier}
}

@article{mazaheri2013image,
  title={Image artifacts on prostate diffusion-weighted magnetic resonance imaging: trade-offs at 1.5 Tesla and 3.0 Tesla},
  author={Mazaheri, Yousef and Vargas, H Alberto and Nyman, Gregory and Akin, Oguz and Hricak, Hedvig},
  journal={Academic radiology},
  volume={20},
  number={8},
  pages={1041--1047},
  year={2013},
  publisher={Elsevier}
}

@article{tamada2021diffusion,
  title={Diffusion-weighted imaging in prostate cancer},
  author={Tamada, Tsutomu and Ueda, Yu and Ueno, Yoshiko and Kojima, Yuichi and Kido, Ayumu and Yamamoto, Akira},
  journal={Magnetic Resonance Materials in Physics, Biology and Medicine},
  pages={1--15},
  year={2021},
  publisher={Springer}
}

@article{jambor2017optimization,
  title={Optimization of prostate MRI acquisition and post-processing protocol: a pictorial review with access to acquisition protocols},
  author={Jambor, Ivan},
  journal={Acta Radiologica Open},
  volume={6},
  number={12},
  pages={2058460117745574},
  year={2017},
  publisher={SAGE Publications Sage UK: London, England}
}

@article{molendowska2024diffusion,
  title={Diffusion MRI in prostate cancer with ultra-strong whole-body gradients},
  author={Molendowska, Malwina and Palombo, Marco and Foley, Kieran G and Narahari, Krishna and Fasano, Fabrizio and Jones, Derek K and Alexander, Daniel C and Panagiotaki, Eleftheria and Tax, Chantal MW},
  journal={NMR in Biomedicine},
  volume={37},
  number={12},
  pages={e5229},
  year={2024},
  publisher={Wiley Online Library}
}

@article{zhang2020prostate,
  title={Prostate diffusion MRI with minimal echo time using eddy current nulled convex optimized diffusion encoding},
  author={Zhang, Zhaohuan and Moulin, Kevin and Aliotta, Eric and Shakeri, Sepideh and Afshari Mirak, Sohrab and Hosseiny, Melina and Raman, Steven and Ennis, Daniel B and Wu, Holden H},
  journal={Journal of Magnetic Resonance Imaging},
  volume={51},
  number={5},
  pages={1526--1539},
  year={2020},
  publisher={Wiley Online Library}
}

@article{zhang2024high,
  title={High-resolution prostate diffusion MRI using eddy current-nulled convex optimized diffusion encoding and random matrix theory-based denoising},
  author={Zhang, Zhaohuan and Aygun, Elif and Shih, Shu-Fu and Raman, Steven S and Sung, Kyunghyun and Wu, Holden H},
  journal={Magnetic Resonance Materials in Physics, Biology and Medicine},
  volume={37},
  number={4},
  pages={603--619},
  year={2024},
  publisher={Springer}
}

@article{kaye2020accelerating,
  title={Accelerating prostate diffusion-weighted MRI using a guided denoising convolutional neural network: retrospective feasibility study},
  author={Kaye, Elena A and Aherne, Emily A and Duzgol, Cihan and H{\"a}ggstr{\"o}m, Ida and Kobler, Erich and Mazaheri, Yousef and Fung, Maggie M and Zhang, Zhigang and Otazo, Ricardo and Vargas, Hebert A and others},
  journal={Radiology: Artificial Intelligence},
  volume={2},
  number={5},
  pages={e200007},
  year={2020},
  publisher={Radiological Society of North America}
}

@article{pfaff2024enhancing,
  title={Enhancing diffusion-weighted prostate MRI through self-supervised denoising and evaluation},
  author={Pfaff, Laura and Darwish, Omar and Wagner, Fabian and Thies, Mareike and Vysotskaya, Nastassia and Hossbach, Julian and Weiland, Elisabeth and Benkert, Thomas and Eichner, Cornelius and Nickel, Dominik and others},
  journal={Scientific Reports},
  volume={14},
  number={1},
  pages={24292},
  year={2024},
  publisher={Nature Publishing Group UK London}
}

@inproceedings{han2023prostate,
  title={Prostate MRI Super-Resolution using Discrete Residual Diffusion Model},
  author={Han, Zhitao and Huang, Wenhui},
  booktitle={2023 IEEE International Conference on Bioinformatics and Biomedicine (BIBM)},
  pages={1947--1950},
  year={2023},
  organization={IEEE}
}

@article{marchenko1967distribution,
  title={Distribution of eigenvalues for some sets of random matrices},
  author={Marchenko, VA and Pastur, Leonid A},
  journal={Mat. Sb.(NS)},
  volume={72},
  number={114},
  pages={507--536},
  year={1967}
}

@article{palombo2023joint,
  title={Joint estimation of relaxation and diffusion tissue parameters for prostate cancer with relaxation-VERDICT MRI},
  author={Palombo, Marco and Valindria, Vanya and Singh, Saurabh and Chiou, Eleni and Giganti, Francesco and Pye, Hayley and Whitaker, Hayley C and Atkinson, David and Punwani, Shonit and Alexander, Daniel C and others},
  journal={Scientific Reports},
  volume={13},
  number={1},
  pages={2999},
  year={2023},
  publisher={Nature Publishing Group UK London}
}

@article{grussu2021feasibility,
  title={Feasibility of data-driven, model-free quantitative MRI protocol design: Application to brain and prostate diffusion-relaxation imaging},
  author={Grussu, Francesco and Blumberg, Stefano B and Battiston, Marco and Kakkar, Lebina S and Lin, Hongxiang and Ianu{\c{s}}, Andrada and Schneider, Torben and Singh, Saurabh and Bourne, Roger and Punwani, Shonit and others},
  journal={Frontiers in Physics},
  volume={9},
  pages={752208},
  year={2021},
  publisher={Frontiers Media SA}
}

@article{ueno2022diffusion,
  title={Diffusion and quantification of diffusion of prostate cancer},
  author={Ueno, Yoshiko and Tamada, Tsutomu and Sofue, Keitaro and Murakami, Takamichi},
  journal={The British Journal of Radiology},
  volume={95},
  number={1131},
  pages={20210653},
  year={2022},
  publisher={The British Institute of Radiology.}
}

@article{buerger2015comparing,
  title={Comparing nonrigid registration techniques for motion corrected MR prostate diffusion imaging},
  author={Buerger, C and S{\'e}n{\'e}gas, J and Kabus, S and Carolus, H and Schulz, H and Agarwal, H and Turkbey, B and Choyke, PL and Renisch, S},
  journal={Medical physics},
  volume={42},
  number={1},
  pages={69--80},
  year={2015},
  publisher={Wiley Online Library}
}

@article{kabus2010fast,
  title={Fast elastic image registration},
  author={Kabus, Sven and Lorenz, Cristian},
  journal={Medical Image Analysis for the Clinic: A Grand Challenge},
  volume={89},
  pages={81--89},
  year={2010}
}

@article{usman2020joint,
  title={Joint B0 and image estimation integrated with model based reconstruction for field map update and distortion correction in prostate diffusion MRI},
  author={Usman, Muhammad and Kakkar, Lebina and Matakos, Antonis and Kirkham, Alex and Arridge, Simon and Atkinson, David},
  journal={Magnetic Resonance Imaging},
  volume={65},
  pages={90--99},
  year={2020},
  publisher={Elsevier}
}

@article{usman2019model,
  title={Model-based reconstruction framework for correction of signal pile-up and geometric distortions in prostate diffusion MRI},
  author={Usman, Muhammad and Kakkar, Lebina and Kirkham, Alex and Arridge, Simon and Atkinson, David},
  journal={Magnetic Resonance in Medicine},
  volume={81},
  number={3},
  pages={1979--1992},
  year={2019},
  publisher={Wiley Online Library}
}

@article{tong2019exploratory,
  title={Exploratory study of geometric distortion correction of prostate diffusion-weighted imaging using B0 map acquisition},
  author={Tong, Angela and Lemberskiy, Gregory and Huang, Chenchan and Shanbhogue, Krishna and Feiweier, Thorsten and Rosenkrantz, Andrew B},
  journal={Journal of Magnetic Resonance Imaging},
  volume={50},
  number={5},
  pages={1614--1619},
  year={2019},
  publisher={Wiley Online Library}
}

@mastersthesis{syversen2018implementation,
  title={Implementation of geometric distortion correction of EPI images in clinical workflow},
  author={Syversen, Ingrid Framås},
  school={NTNU},
  year={2018}
}

@article{rakowpenner2015prostate,
  title={Prostate diffusion imaging with distortion correction},
  author={Rakow-Penner, Rebecca A and White, Nathan S and Margolis, Daniel JA and Parsons, John Kellogg and Schenker-Ahmed, Natalie and Kuperman, Joshua M and Bartsch, Hauke and Choi, Hyung W and Bradley, William G and Shabaik, Ahmed and Huang, Jiaoti and Liss, Michael A and Marks, Leonard and Kane, Christopher J and Reiter, Robert E and Raman, Steven S and Karow, David S and Dale, Anders M},
  journal={Magnetic Resonance Imaging},
  volume={33},
  number={9},
  pages={1178--1181},
  year={2015},
  publisher={Elsevier}
}

@article{nketiah2018geometric,
  title={Geometric distortion correction in prostate diffusion-weighted MRI and its effect on quantitative apparent diffusion coefficient analysis},
  author={Nketiah, Gabriel and Selnæs, Kirsten M and Sandsmark, Elise and Teruel, Jose R and Krüger-Stokke, Brage and Bertilsson, Helena and Bathen, Tone F and Elschot, Mattijs},
  journal={Magnetic Resonance in Medicine},
  volume={79},
  number={5},
  pages={2524--2532},
  year={2018},
  publisher={Wiley Online Library}
}

@article{ciris2019accelerated,
  title={Accelerated Segmented Diffusion-Weighted Prostate Imaging for Higher Resolution, Higher Geometric Fidelity and Multi-b Perfusion Estimation},
  author={Ciris, Pelin Aksit and Chiou, Jr-yuan George and Glazer, Daniel and Chao, Tzu-Cheng and Tempany-Afdhal, Clare M and Madore, Bruno and Maier, Stephan E},
  journal={Investigative Radiology},
  volume={54},
  number={4},
  pages={238--246},
  year={2019},
  publisher={Lippincott Williams \& Wilkins}
}

@article{bergen2020field,
  title={Field-map correction in read-out segmented echo planar imaging for reduced spatial distortion in prostate DWI for MRI-guided radiotherapy applications},
  author={Bergen, Robert V and Ryner, Lawrence and Essig, Marco},
  journal={Magnetic Resonance Imaging},
  volume={67},
  pages={43--49},
  year={2020},
  publisher={Elsevier}
}

@article{lemberskiy2018characterization,
  title={Characterization of prostate microstructure using water diffusion and NMR relaxation},
  author={Lemberskiy, Gregory and Fieremans, Els and Veraart, Jelle and Deng, Fang-Ming and Rosenkrantz, Andrew B and Novikov, Dmitry S},
  journal={Frontiers in physics},
  volume={6},
  pages={91},
  year={2018},
  publisher={Frontiers Media SA}
}

@article{andersson2016integrated,
  title={An integrated approach to correction for off-resonance effects and subject movement in diffusion MR imaging},
  author={Andersson, Jesper LR and Sotiropoulos, Stamatios N},
  journal={Neuroimage},
  volume={125},
  pages={1063--1078},
  year={2016},
  publisher={Elsevier}
}

@article{godley2018accuracy,
  title={Accuracy of high b-value diffusion-weighted MRI for prostate cancer detection: a meta-analysis},
  author={Godley, Keith Craig and Syer, Tom Joseph and Toms, Andoni Paul and Smith, Toby Oliver and Johnson, Glyn and Cameron, Donnie and Malcolm, Paul Napier},
  journal={Acta Radiologica},
  volume={59},
  number={1},
  pages={105--113},
  year={2018},
  publisher={Sage Publications Sage UK: London, England}
}

@inproceedings{lin2017focal,
  title={Focal loss for dense object detection},
  author={Lin, Tsung-Yi and Goyal, Priya and Girshick, Ross and He, Kaiming and Doll{\'a}r, Piotr},
  booktitle={Proceedings of the IEEE international conference on computer vision},
  pages={2980--2988},
  year={2017}
}

@article{shanmugam2020and,
  title={When and why test-time augmentation works},
  author={Shanmugam, Divya and Blalock, Davis and Balakrishnan, Guha and Guttag, John},
  journal={arXiv preprint arXiv:2011.11156},
  volume={1},
  number={3},
  pages={4},
  year={2020}
}

@article{goldenberg2019new,
  title={A new era: artificial intelligence and machine learning in prostate cancer},
  author={Goldenberg, S Larry and Nir, Guy and Salcudean, Septimiu E},
  journal={Nature Reviews Urology},
  volume={16},
  number={7},
  pages={391--403},
  year={2019},
  publisher={Nature Publishing Group UK London}
}

@article{van2021systematic,
  title={A systematic review of artificial intelligence in prostate cancer},
  author={Van Booven, Derek J and Kuchakulla, Manish and Pai, Raghav and Frech, Fabio S and Ramasahayam, Reshna and Reddy, Pritika and Parmar, Madhumita and Ramasamy, Ranjith and Arora, Himanshu},
  journal={Research and reports in urology},
  pages={31--39},
  year={2021},
  publisher={Taylor \& Francis}
}

@article{schelb2019classification,
  title={Classification of cancer at prostate MRI: deep learning versus clinical PI-RADS assessment},
  author={Schelb, Patrick and Kohl, Simon and Radtke, Jan Philipp and Wiesenfarth, Manuel and Kickingereder, Philipp and Bickelhaupt, Sebastian and Kuder, Tristan Anselm and Stenzinger, Albrecht and Hohenfellner, Markus and Schlemmer, Heinz-Peter and others},
  journal={Radiology},
  volume={293},
  number={3},
  pages={607--617},
  year={2019},
  publisher={Radiological Society of North America}
}

@article{nachar2008mann,
  title={The Mann-Whitney U: A test for assessing whether two independent samples come from the same distribution},
  author={Nachar, Nadim},
  journal={Tutorials in Quantitative Methods for Psychology},
  volume={4},
  number={1},
  pages={13--20},
  year={2008}
}

@article{kang2020balanced,
  title={Balanced training/test set sampling for proper evaluation of classification models},
  author={Kang, Donghoon and Oh, Sejong},
  journal={Intelligent Data Analysis},
  volume={24},
  number={1},
  pages={5--18},
  year={2020},
  publisher={SAGE Publications Sage UK: London, England}
}

@article{grandini2020metrics,
  title={Metrics for multi-class classification: an overview},
  author={Grandini, Margherita and Bagli, Enrico and Visani, Giorgio},
  journal={arXiv preprint arXiv:2008.05756},
  year={2020}
}

@article{yildirim2022deep,
  title={Deep learning-based PI-RADS score estimation to detect prostate cancer using multiparametric magnetic resonance imaging},
  author={Yildirim, Kadir and Yildirim, Muhammed and Eryesil, Hasan and Talo, Muhammed and Yildirim, Ozal and Karabatak, Murat and Ogras, Mehmet Sezai and Artas, Hakan and Acharya, U Rajendra},
  journal={Computers and Electrical Engineering},
  volume={102},
  pages={108275},
  year={2022},
  publisher={Elsevier}
}

@article{youn2021detection,
  title={Detection and PI-RADS classification of focal lesions in prostate MRI: Performance comparison between a deep learning-based algorithm (DLA) and radiologists with various levels of experience},
  author={Youn, Seo Yeon and Choi, Moon Hyung and Kim, Dong Hwan and Lee, Young Joon and Huisman, Henkjan and Johnson, Evan and Penzkofer, Tobias and Shabunin, Ivan and Winkel, David Jean and Xing, Pengyi and others},
  journal={European Journal of Radiology},
  volume={142},
  pages={109894},
  year={2021},
  publisher={Elsevier}
}

@article{liu2021textured,
  title={Textured-based deep learning in prostate cancer classification with 3T multiparametric MRI: comparison with PI-RADS-based classification},
  author={Liu, Yongkai and Zheng, Haoxin and Liang, Zhengrong and Miao, Qi and Brisbane, Wayne G and Marks, Leonard S and Raman, Steven S and Reiter, Robert E and Yang, Guang and Sung, Kyunghyun},
  journal={Diagnostics},
  volume={11},
  number={10},
  pages={1785},
  year={2021},
  publisher={MDPI}
}

@article{yoo2019prostate,
  title={Prostate cancer detection using deep convolutional neural networks},
  author={Yoo, Sunghwan and Gujrathi, Isha and Haider, Masoom A and Khalvati, Farzad},
  journal={Scientific reports},
  volume={9},
  number={1},
  pages={19518},
  year={2019},
  publisher={Nature Publishing Group UK London}
}

@article{winkel2020autonomous,
  title={Autonomous detection and classification of PI-RADS lesions in an MRI screening population incorporating multicenter-labeled deep learning and biparametric imaging: proof of concept},
  author={Winkel, David J and Wetterauer, Christian and Matthias, Marc Oliver and Lou, Bin and Shi, Bibo and Kamen, Ali and Comaniciu, Dorin and Seifert, Hans-Helge and Rentsch, Cyrill A and Boll, Daniel T},
  journal={Diagnostics},
  volume={10},
  number={11},
  pages={951},
  year={2020},
  publisher={MDPI}
}

@article{scikit-learn,
  title={Scikit-learn: Machine Learning in {P}ython},
  author={Pedregosa, F. and Varoquaux, G. and Gramfort, A. and Michel, V.
          and Thirion, B. and Grisel, O. and Blondel, M. and Prettenhofer, P.
          and Weiss, R. and Dubourg, V. and Vanderplas, J. and Passos, A. and
          Cournapeau, D. and Brucher, M. and Perrot, M. and Duchesnay, E.},
  journal={Journal of Machine Learning Research},
  volume={12},
  pages={2825--2830},
  year={2011}
}

@article{richardson2024receiver,
  title={The receiver operating characteristic curve accurately assesses imbalanced datasets},
  author={Richardson, Eve and Trevizani, Raphael and Greenbaum, Jason A and Carter, Hannah and Nielsen, Morten and Peters, Bjoern},
  journal={Patterns},
  volume={5},
  number={6},
  year={2024},
  publisher={Elsevier}
}

@inproceedings{davis2006relationship,
  title={The relationship between Precision-Recall and ROC curves},
  author={Davis, Jesse and Goadrich, Mark},
  booktitle={Proceedings of the 23rd international conference on Machine learning},
  pages={233--240},
  year={2006}
}

@article{buckland1994relationship,
  title={The relationship between recall and precision},
  author={Buckland, Michael and Gey, Fredric},
  journal={Journal of the American society for information science},
  volume={45},
  number={1},
  pages={12--19},
  year={1994},
  publisher={Wiley Online Library}
}

@inproceedings{guo2017calibration,
    title     = {On Calibration of Modern Neural Networks},
    author    = {Guo, Chuan and Pleiss, Geoff and Sun, Yu and Weinberger, Kilian Q.},
    booktitle = {Proceedings of the 34th International Conference on Machine Learning},
    series    = {Proceedings of Machine Learning Research},
    volume    = {70},
    pages     = {1321--1330},
    year      = {2017},
    publisher = {PMLR},
    url       = {https://proceedings.mlr.press/v70/guo17a.html},
    eprint    = {1706.04599},
    archivePrefix = {arXiv}
  }

@article{irfanoglu2015dr,
  title={DR-BUDDI (Diffeomorphic Registration for Blip-Up blip-Down Diffusion Imaging) method for correcting echo planar imaging distortions},
  author={Irfanoglu, M Okan and Modi, Pooja and Nayak, Amritha and Hutchinson, Elizabeth B and Sarlls, Joelle and Pierpaoli, Carlo},
  journal={Neuroimage},
  volume={106},
  pages={284--299},
  year={2015},
  publisher={Elsevier}
}

@inproceedings{haber2006intensity,
  author    = {Haber, Eldad and Modersitzki, Jan},
  title     = {Intensity Gradient Based Registration and Fusion of Multi-modal Images},
  booktitle = {Medical Image Computing and Computer-Assisted Intervention (MICCAI) 2006},
  series    = {Lecture Notes in Computer Science},
  volume    = {4191},
  pages     = {726--733},
  year      = {2006},
  publisher = {Springer},
  doi       = {10.1007/11866763_89}
}

@article{maes1997multimodality,
  author  = {Maes, Frederik and Collignon, Andr{\'e} and Vandermeulen, Dirk and Marchal, Guy and Suetens, Paul},
  title   = {Multimodality Image Registration by Maximization of Mutual Information},
  journal = {IEEE Transactions on Medical Imaging},
  volume  = {16},
  number  = {2},
  pages   = {187--198},
  year    = {1997},
  doi     = {10.1109/42.563664}
}
